# Harnessing optical micro-combs for microwave photonics

Jiayang Wu,[1,†] Xingyuan Xu,[1,†] Thach G. Nguyen,[2] Sai T. Chu,[3] Brent E. Little,[4] Roberto Morandotti,[5,6,7] Arnan Mitchell,[2] and David J. Moss[1]*

[1]*Centre for Micro-Photonics, Swinburne University of Technology, Hawthorn, VIC 3122, Australia*
[2]*ARC Centre of Excellence for Ultrahigh-bandwidth Devices for Optical Systems (CUDOS), RMIT University, Melbourne, VIC 3001, Australia*
[3]*Department of Physics and Material Science, City University of Hong Kong, Tat Chee Avenue, Hong Kong, China.*
[4]*State Key Laboratory of Transient Optics and Photonics, Xi'an Institute of Optics and Precision Mechanics, Chinese Academy of Science, Xi'an, China.*
[5]*INSR-Énergie, Matériaux et Télécommunications, 1650 Boulevard Lionel-Boulet, Varennes, Québec, J3X 1S2, Canada.*
[6]*National Research University of Information Technologies, Mechanics and Optics, St. Petersburg, Russia.*
[7]*Institute of Fundamental and Frontier Sciences, University of Electronic Science and Technology of China, Chengdu 610054, China.*

[†]*These authors contribute equally to this paper.*
*\*dmoss@swin.edu.au*

## Abstract

In the past decade, optical frequency combs generated by high-Q micro-resonators, or micro-combs, which feature compact device footprints, high energy efficiency, and high-repetition-rates in broad optical bandwidths, have led to a revolution in a wide range of fields including metrology, mode-locked lasers, telecommunications, RF photonics, spectroscopy, sensing, and quantum optics. Among these, an application that has attracted great interest is the use of micro-combs for RF photonics, where they offer enhanced functionalities as well as reduced size and power consumption over other approaches. This article reviews the recent advances in this emerging field. We provide an overview of the main achievements that have been obtained to date, and highlight the strong potential of micro-combs for RF photonics applications. We also discuss some of the open challenges and limitations that need to be met for practical applications.

## I. INTRODUCTION

Radio-frequency (RF) photonics, which merges photonic technologies with microwave or millimeter wave (MMW) engineering, has been widely used for many applications in telecommunications, radar, metrology, sensing, testing and imaging [1-3]. The history of RF photonics can be traced back to the 1970's [4]. In parallel with fibre optics advances, the development of techniques for the transmission and processing of wide-bandwidth analogue RF signals using photonics technologies has been very active and fast-growing in the last few decades. These two fields complement each other and offer revolutionary new technologies for the information age.

Optical frequency combs (OFCs), which are often referred to as optical rulers since their spectra consist of a precise sequence of discrete and equally spaced spectral lines that represent precise marks in frequency, has revolutionized the fields of optical metrology and synthesis since the first demonstration in 2000 [5-7]. Fully stabilized OFCs solved the challenge of directly measuring optical frequencies and are now exploited as the most accurate time references, ready to replace the current standard for time [8]. Their importance was recognized worldwide with the 2005 Nobel Prize awarded to T. W. Hänsch and J. Hall for their breakthroughs [9]. Since then, the applications of OFCs have been extended to a much wider scope, from spectroscopy [10-13], optical clocks [14, 15], to quantum optics [16-18], telecommunications [19, 20], and RF photonics [21-23]. However, increasing challenges have emerged as the specific requirements in these expanding applications become more demanding and cannot be fully met by OFC generation methods based on mode-locked Ti:Sapphire or fibre lasers. In RF photonics, for instance, both high repetition rates (>10 GHz) and flexibility in tuning independently the repetition rates and central frequencies are required for applications to arbitrary waveform generation [21, 24], but these requirements cannot easily be met by solid state or fibre lasers.

Recent advancements in the fabrication of optical micro-resonators with ultrahigh quality (Q) factors have opened up promising new avenues towards realising new types of OFC sources. OFCs generated by high-Q micro-resonators, or so-called "micro-combs", typically feature compact device footprints, high energy efficiency, and high-repetition-rates in broad optical bandwidths, which offer new possibilities in a wide range of fields such as metrology [25-27], mode-locked lasers [28-31], telecommunications [32-35], RF photonics [36-39], and quantum optics [40, 41]. While micro-combs have been the subject of several reviews [42-47], here we focus on their role in RF photonics. More specifically, we use the term "RF photonics" to delineate a subset of photonic technologies that is utilized to detect, distribute, process, frequency translate, and switch optical signals in the RF, microwave, and MMW frequency regions. We review exciting new developments in the field of micro-comb based RF photonics, with a perspective on both their strong potential as well as on open questions and challenges that still remain.

This paper is structured as follows. The generation of micro-combs, in terms of modelling and principles, as well as devices and their fabrication, is presented in Section II. In Section III, we review recent work in RF photonics based micro-combs, including RF photonic oscillators for microwave and MMW generation, optical true time delay lines for phased array antennas, RF photonic filters, and multi-functional RF photonic processors. The open challenges and limitations of RF photonics based on micro-combs are discussed in Section IV, followed by conclusions in Section V.

## II. OPTICAL MICRO-COMBS

### 1. Micro-comb platforms

Optical micro-combs were first demonstrated in 2007 [48], while the first reports based on photonic integrated platforms followed in 2010 [49-52]. To date, dielectric micro-resonators made from a variety of materials, platforms, and structures for the generation of micro-combs have been reported [51]. Figure 1 shows a range of micro-resonator platforms for which micro-comb generation has been reported, ranging from silica, magnesium and calcium fluorides ($MgF_2$ and $CaF_2$), to silicon, silicon nitride ($Si_3N_4$), high-index doped silica (Hydex) glass, aluminium nitride (AlN), and diamond. The device structures include bulk and monolithic toroids, spheres, as well as integrated ring resonators. Most of the devices are based on whispering gallery mode (WGM) [53, 54] resonances and have the common feature of possessing exceptionally high Q factors. Generally speaking, the ideal micro-comb must be well tailored to a particular application, and devices based on different materials and architectures each have their advantages for micro-comb generation and/or are better suited in specific applications. Table 1 compares the key parameters of a range of devices for micro-comb generation from the visible to mid-infrared (MIR) wavelengths.

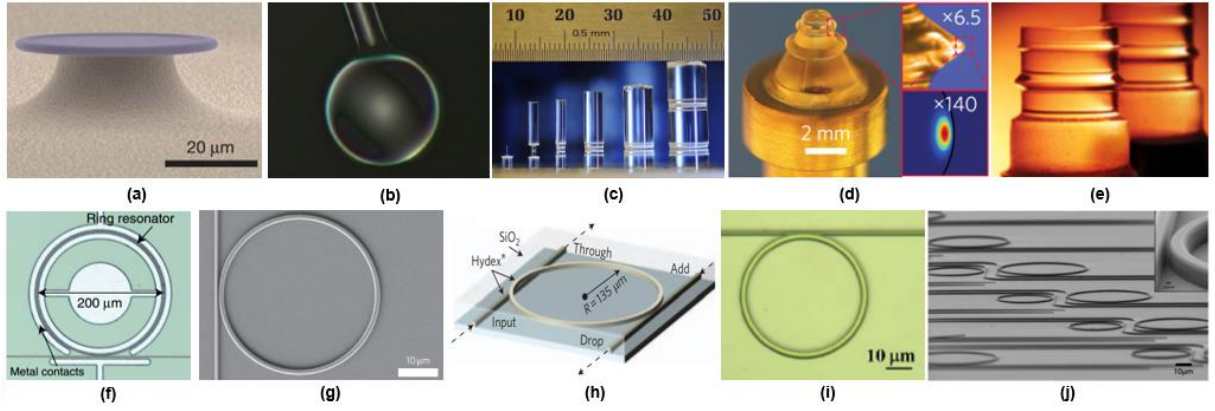

Fig. 1. Dielectric micro-resonators made from a variety of materials and various structures used for the generation of micro-combs. (a) silica toroid [48]. (b) silica sphere [55]. (c) silica rod [56]. (d) magnesium fluorides toroid [57]. (e) calcium fluorides toroid [58]. (f) silicon microring resonator [59]. (g) silicon nitride microring resonator [50]. (h) Hydex microring resonator [49]. (i) aluminium nitride microring resonator [60]. (j) diamond microring resonator [61].

**TABLE I   KEY PARAMETERS OF DEVICES USED FOR MICRO-COMB GENERATION**

| Material | $n_2$* ($m^2W^{-1}$) | Structure | Q factor | Comb spacing | Wavelength range | References |
|---|---|---|---|---|---|---|
| $SiO_2$ | $2.6\times10^{-20}$ | toroid | $10^8$ | ~375 GHz | 1200 ~ 1700 nm | [48] |
| $SiO_2$ | $2.6\times10^{-20}$ | toroid | $2\times10^8$ | ~375 GHz | 990 ~ 2170 nm | [36] |
| $SiO_2$ | $2.6\times10^{-20}$ | sphere | $2\times10^7$ | ~427 GHz | 1450 ~ 1700 nm | [55] |
| $SiO_2$ | $2.6\times10^{-20}$ | rod | $5\times10^8$ | ~32.6 GHz | 1510 ~ 1610 nm | [56] |
| $MgF_2$ | $1\times10^{-20}$ | toroid | ~$10^9$ | ~35 GHz | 1533 ~ 1553 nm | [62] |
| $MgF_2$ | $1\times10^{-20}$ | toroid | >$10^9$ | 10−110 GHz | 2350 ~ 2550 nm | [63] |
| $MgF_2$ | $1\times10^{-20}$ | toroid | ~$10^{10}$ | Non-uniform | 0.36 ~ 1.6 µm | [64] |
| $CaF_2$ | $3.2\times10^{-20}$ | truncated sphere | $6\times10^9$ | ~13 GHz | 1545 nm ~ 1575 nm | [65] |
| $CaF_2$ | $3.2\times10^{-20}$ | truncated sphere | $3\times10^9$ | ~23.78 GHz | near 794 nm | [66] |
| Hydex | $1.15\times10^{-19}$ | ring | $1.2\times10^6$ | 0.2−6 THz | 1400 ~ 1700 nm | [50] |
| Hydex | $1.15\times10^{-19}$ | ring | $1.5\times10^6$ | ~49 GHz | 1460 ~ 1660 nm | [67] |
| $Si_3N_4$ | $2.5\times10^{-19}$ | ring | $5\times10^5$ | ~403 GHz | 1450 ~ 1750 nm | [50] |
| $Si_3N_4$ | $2.5\times10^{-19}$ | ring | $10^5$ | ~226 GHz | 1170 ~ 2350 nm | [68] |
| $Si_3N_4$ | $2.5\times10^{-19}$ | ring | $2.6\times10^5$ | ~977.2 GHz | 502 ~ 580 nm | [69] |
| $Si_3N_4$ | $2.5\times10^{-19}$ | ring | $1.7\times10^7$ | ~25 GHz | 1510 ~ 1600 nm | [70] |
| Si | $6\times10^{-18}$ | ring | $5.9\times10^5$ | ~127 GHz | 2.1 ~ 3.5 µm | [59] |
| AlN | $(2.3\pm1.5)\times10^{-19}$ | ring | $6\times10^5$ | ~370 GHz | 1450 ~ 1650 nm | [71] |
| AlN | $(2.3\pm1.5)\times10^{-19}$ | ring | $6\times10^5$ | ~369 GHz | near 517 & 776 nm | [60] |
| Diamond | $(8.2\pm3.5)\times10^{-20}$ | ring | ~$10^6$ | ~ 925 GHz | 1516 ~ 1681 nm | [61] |

* The Kerr coefficient $n_2$ of different materials here are the measured values around 1550 nm, detailed information can be seen in Refs.[50, 52, 61, 63, 71-74].

Silica micro-toroids (Fig. 1(a)) were used in the first demonstration of optical micro-combs [48]. Although silica's Kerr nonlinearity is not high ($n_2 = \sim 2.6 \times 10^{-20}$ m$^2$W$^{-1}$) [52], its extremely low material absorption – both linear and nonlinear – over a broad wavelength range from the visible to the MIR allows for extremely high Q factors of up to $5 \times 10^8$ [75, 76]. Micro-spheres (Fig. 1(b)) can achieve even higher Q factors up to $10^{10}$ [77, 78], although they have many more degenerate modes, which is undesirable for controlling mode crossing – an important parameter in comb formation [36, 48, 79]. Micro-rod resonators (Fig. 1(c)) have Q factors comparable to micro-toroids, and are typically fabricated by using a $CO_2$ laser to shape and polish silica rods [56].

The transparency window of crystalline materials such as $MgF_2$ and $CaF_2$ extends from the ultraviolet (~160 nm) to the MIR (~7 μm), which is even broader than fused silica, where the absorption increases significantly beyond 2.2 μm [63]. Figures 1(d) and (e) show the micro-toroids made from $MgF_2$ and $CaF_2$, respectively. For wavelengths longer than 1440 nm, $MgF_2$ has anomalous group velocity dispersion, which facilitates phase matching and reduces the threshold of hyper-parametric oscillation for comb generation. The relatively small thermo-refractive coefficient of $MgF_2$ also increases the threshold of the thermo-optical instabilities and the stability of phase locking. The thermo-refractive and thermal-expansion coefficients of $MgF_2$ are both positive at room temperature, which allows the cavity mode to be thermally self-locked and enables the stabilization of the micro-comb repetition rate and offset frequency [38, 62-64]. On the contrary, materials with negative coefficients, such as $CaF_2$, suffer from high thermal oscillatory instabilities and usually require active mode-locking schemes [63]. Since thermal reflow cannot be used for crystalline materials, grinding and polishing techniques are employed to obtain the smooth surfaces [80, 81] necessary to achieve Q factors as high as $10^{11}$ in $CaF_2$ resonators [82]. Recently, other metal difluorides such as barium fluoride ($BaF_2$) and strontium fluoride ($SrF_2$) have also been investigated for ultrahigh-Q resonators as well as micro-comb generation [81, 83, 84].

Silicon is the foundation of the integrated photonics industry, largely because of its synergy with computer chip technology (complementary metal–oxide–semiconductor (CMOS)) [85]. Its refractive index ($n_{Si} = \sim 3.48$), higher than silica ($n_{SiO2} = \sim 1.44$) or silicon nitride ($n_{Si3N4} = \sim 1.98$), allows tight mode confinement within compact device footprints of 10's to 100's of microns [86-88], which, when combined with a high Kerr coefficient ($n_2 = \sim 6 \times 10^{-18}$ m$^2$ W$^{-1}$), yields extremely high nonlinear parameters ($\gamma$) of 300,000 W$^{-1}$ km$^{-1}$ [74]. However, silicon suffers from strong two photon absorption (TPA) in the telecommunications band, leading to a very low nonlinear figure of merit (FOM = $n_2/(\beta_{TPA} \times \lambda)$, where $\beta_{TPA}$ is the TPA coefficient and $\lambda$ is the wavelength) of only ~0.3 [52]. This limit intrinsically arises from the silicon's band structure, and cannot be compensated for by engineering waveguide dimensions. In the MIR wavelength range, however, silicon is transparent to both one- and two-photon transitions, and thus it is a highly attractive platform for nonlinear photonics in this band. Figure. 3(f) shows a silicon microring resonator (MRR) used for MIR micro-comb generation [59], where a high Q factor of $5.9 \times 10^5$ was achieved by using etchless fabrication methods, and a p-i-n junction was employed to sweep out the free carriers generated by three (and higher) photon absorption.

The low FOM of silicon in the telecommunication band was the motivation for developing alternate CMOS-compatible platforms for nonlinear optics, including silicon nitride, Hydex glass, amorphous silicon, silicon-rich nitride, and so forth [52]. Among them, silicon nitride and Hydex have been widely used for on-chip generation of optical micro-combs.

Silicon nitride ($Si_3N_4$), a CMOS-compatible material widely used as a dielectric insulator in computer chips, was proposed as a platform for nonlinear integrated optics in 2007 [89]. Its Kerr coefficient ($n_2 = \sim 2.5 \times 10^{-19}$ m$^2$

W$^{-1}$ [50]) is an order of magnitude higher than fused silica while the TPA coefficient ($\beta_{TPA}$) is negligible in the telecommunications band, thus yielding a FOM >>1 [52]. However, its tensile film stress makes it challenging to grow layers thicker than 400 nm without cracks [90]; this is a problem since much thicker layers are needed for high mode confinement and for phase matching via dispersion engineering [90, 91]. Nonetheless, in 2010, Si$_3$N$_4$ MRRs over 700-nm thick (Fig. 3(g)) and with Q factors of the order of 10$^5$ were realised for micro-comb generation [50]. They were fabricated by using a thermal cycling process of low-pressure chemical vapour deposition (LPCVD). To control film stress and prevent cracking, other approaches such as introducing mechanical trenches [90] or using a photonic damascene process based on substrate topography [91] have also been proposed and used to achieve Si$_3$N$_4$ film thicknesses of 910 nm and 1.35 μm, respectively. By exploiting mode interactions between coupled Si$_3$N$_4$ MRRs, Kerr frequency combs have been generated even in 300-nm-thick Si$_3$N$_4$ devices with normal dispersion [44, 92-96]. Also, nonlinear processes such as second-harmonic generation (SHG), sum-frequency generation (SFG), and third-harmonic generation (THG) have been utilized to extend Si$_3$N$_4$ micro-combs into the visible range where the large normal material dispersion typically dominates [69, 96, 97].

Hydex was developed as a CMOS compatible optical platform for linear optics starting in the early 2000's [98] and applied to nonlinear optics in 2008 [99]. Its success as a nonlinear optical platform [31, 40, 49, 52, 99, 100] can be attributed to its ultralow linear loss (~5−7 dB·m$^{-1}$), a sufficiently higj nonlinear parameter ($\gamma$ = ~233 W$^{-1}$·km$^{-1}$, 200 times that of standard single-mode telecommunication fibres), and in particular a negligible nonlinear loss up to extremely high intensities (~25 GW·cm$^{-2}$). Its refractive index ($n$ = ~1.5−1.9) is slightly lower than that of Si$_3$N$_4$ and considerably lower than Si, and so buried waveguide geometries are typically used rather than nanowires. Nonetheless, the core-cladding contrast (typically ~17%) is still high enough to allow relatively tight bend radii down to ~20 μm. Its Kerr coefficient is ~1.15 × 10$^{-19}$ m$^2$ W$^{-1}$ [52], or about 5x silica. Figure 3(h) shows a schematic of a Hydex MRR used for micro-comb generation, with a Q factor of 1.2 × 10$^6$ [49].

Other integrated platforms, such as aluminium nitride (AlN) and diamond, are not CMOS compatible but are still powerfully suited for micro-comb generation. Aluminium nitride (AlN), with a Kerr coefficient ((2.3 ± 1.5) × 10$^{-19}$ m$^2$ W$^{-1}$ [71]) comparable to Si$_3$N$_4$ and Hydex, is non-centrosymmetric and so has both second- and third-order optical nonlinearities ($\chi^{(2)}$ and $\chi^{(3)}$) that have been conveniently used for micro-comb generation at both near infrared (NIR) and visible wavelengths [60, 71]. Its high thermal conductivity (285 W/m·K, about 100 times higher than Si$_3$N$_4$ [71]) greatly improves heat dissipation and enhances the robustness of micro-combs at high optical powers. Figure 3(i) shows an AlN MRR used for micro-comb generation in the visible wavelengths [60], with a Q factor of ~600,000. Diamond, first proposed as a platform for micro-comb generation in 2014 [61], has a relatively high refractive index ($n$ = ~2.4) and low absorption loss from the ultraviolet to the far-infrared (FIR), with a Kerr nonlinearity of $n_2$ = (8.2 ± 3.5) × 10$^{-20}$ m$^2$ W$^{-1}$ in the telecommunications band. Its large bandgap of 5.5 eV make it exempt from TPA not only in the NIR but also at visible wavelengths. Its high thermal conductivity and low thermo-optic coefficient also enable high power handling capabilities and result in micro-combs which are temperature-stabilized over a wide wavelength range. Figure 3(j) shows a diamond MRR used for micro-comb generation [61], with a Q factor of ~10$^6$.

## 2. Modeling and principles

Figure 2 shows a schematic illustration of micro-comb generation. The physics underlying the generation of micro-combs is the same for all platforms, and is based on optical parametric oscillation (OPO) [101, 102], with parametric gain provided by the Kerr nonlinearity $\chi^{(3)}$. Like other types of oscillators [103], there needs to be enough gain in the resonant cavity to overcome the cavity loss so that oscillation can occur. The entire OFC generation process can be divided into two stages: the first is hyper-parametric oscillation [49], also referred to as modulation instability (MI) in fibres [104], related to the generation of initial new frequency components in a resonant cavity that contains only a pump and low-level noise. When a nonlinear cavity is resonantly excited by a high-power continuous-wave (CW) pump light, two pump photons at a frequency of $\omega_p$ annihilate, generating a signal and idler photon at frequencies of $\omega_s$ and $\omega_i$, respectively. Energy conservation yields:

$$2\omega_p = \omega_s + \omega_i . \tag{1}$$

This process amplifies the noise parametrically (with increasing pump power) until the round-trip gain exceeds the loss, at which point hyper-parametric oscillation occurs. Unlike cascaded four-wave mixing (FWM), hyper-parametric oscillation occurs at the micro-cavity resonance nearest to the parametric gain maximum. Generally, this is not the resonance adjacent to the pump, but rather is determined by the MI gain peak, which depends on a series of factors including dispersion, pump wavelength, pump power, Kerr nonlinearity, as well as linear and nonlinear losses. Once hyper-parametric oscillation occurs, both the lasing wavelength and the pump power in the resonator become clamped. Then in the second stage, by further increasing the pump power, other oscillating modes can be amplified and a process of cascaded FWM can be initiated to generate a series of wavelengths at multiples of this spacing. Rich nonlinear dynamics occur in the second stage, including chaos, breathing, sub-comb formation, and mode-locking transitions [46, 51, 57, 72, 105-107].

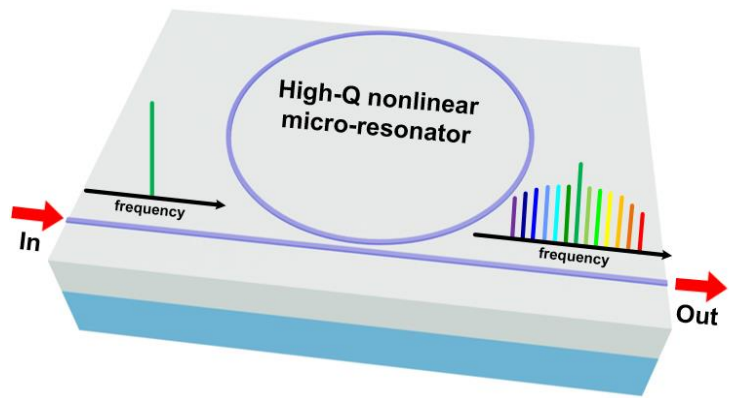

Fig. 2. Schematic illustration of micro-comb generation. A high-power pump is converted into a broadband OFC after going through a high-Q nonlinear micro-resonator.

The modeling of Kerr micro-comb dynamics can be dealt with in two ways – either by spectro-temporal approaches using coupled-mode equations (CMEs) [102, 108], or by spatio-temporal approaches using the Lugiato-Lefever equations (LLEs) [109-112]. Here, we focus on the latter, which have been widely used for the modeling and analyses of comb dynamics such as MI, chaos, and temporal cavity solitons.

The derivation of the LLE starts from the Ikeda map [113] that involves the nonlinear Schrödinger (NLS) equation, which describes the evolution of the slowly varying envelope of the electric field over consecutive round-trips in the resonator. The Ikeda map can be obtained by considering the evolution of the field during each round-trip in the cavity, which requires a partitioning of the temporal duration of the electric field so that it coincides with the round-trip time. Assuming a single spatial mode and monochromatic driving, the relations between the fields at different round trips can be described by [46, 110]:

$$E^{(m+1)}(0,\tau) = \sqrt{1-\kappa}E_{in} + \sqrt{\kappa}E^{(m)}(0,\tau)\exp(i\theta_0), \tag{2}$$

$$\frac{\partial E(z,\tau)}{\partial z} = -\frac{\alpha_i}{2}E + i\sum_{k\geq 2}\frac{\beta_k}{k!}\left(i\frac{\partial}{\partial \tau}\right)^k E + i\gamma|E|^2 E. \tag{3}$$

Here $E^{(m)}(z, \tau)$ is the intracavity field during the $m^{th}$ round-trip, where $z$ is the longitudinal coordinate along the resonator and $\tau$ is time expressed in a reference frame - moving with the group velocity of light at the driving frequency $\omega_0$. $E_{in}$ is the external driving field, and $\kappa$ is the power coupling coefficient of the input/output coupler. $L$ is the round-trip length of the resonant cavity, and $\theta_0$ is the linear phase accumulated over each round-trip. $\alpha_i$ is the linear absorption coefficient of the resonator, $\gamma$ is the nonlinear parameter, and $\beta_k = d^k\beta/d\omega^k|_{\omega=\omega_0}$ are the Taylor series expansion coefficients of the resonator propagation constant $\beta(\omega)$ at $\omega_0$.

For micro-resonators with low cavity loss, the intracavity field exhibits negligible evolution over a single trip. Then, Eqs. (2) and (3) can be averaged into a single mean-field LLE expressed as [110, 114]:

$$t_R \frac{\partial E(t,\tau)}{\partial t} = \left[-\alpha - i\delta_0 + iL\sum_{k\geq 2}\frac{\beta_k}{k!}\left(i\frac{\partial}{\partial \tau}\right)^k + i\gamma|E|^2 L\right]E + \sqrt{1-\kappa}E_{in}. \tag{4}$$

In Eq. (4), $t_R$ is the round-trip time, $\alpha = (\alpha_i L + 1 - \kappa)/2$ is the total cavity loss, and $\delta_0 = 2\pi N - \theta_0$ is the phase detuning of the driving field with respect to the closest linear resonance with an order of $N$. $t$ describes the slow time of the cavity related to the number of round-trips $m$ as $E(t = mt_R, \tau) = E^{(m)}(z = 0, \tau)$. Hence, the field varies with $t$ for timescales longer than the round-trip time $t_R$. By normalizing some of the variables and neglecting high-order dispersion ($\beta_k = 0$ for $k \geq 3$), Eq. (4) can be further written as [44, 115]:

$$\frac{\partial E(t',\tau')}{\partial t'} = \left[-1 + i(|E'|^2 - \Delta) - i\eta\frac{\partial^2}{\partial \tau'^2}\right]E' + S. \tag{5}$$

where $t' = \alpha t/t_R$, $\tau' = \tau\sqrt{2\alpha/(|\beta_2|L)}$, $E' = E\sqrt{\gamma L/\alpha}$, $\Delta = \delta_0 / \alpha$, $d_k = L\beta_k [\sqrt{2\alpha/(|\beta_2|L)}]^k/(\alpha k!)$, $S = E_{in}\sqrt{\gamma L(1-\kappa)/\alpha^3}$, and $\eta = \text{sign}(\beta_2)$. For an input CW pump $E_{in}$, the solutions of Eq. (5) correspond to the CW steady state, at which $\partial E'/\partial \tau' = 0$ and Eq. (5) can be simplified as:

$$[1 + i(\Delta - |E'|^2)]E' = S. \tag{6}$$

For $X = |S|^2$ and $Y = |E'|^2$ that denote the normalized pump and stationary intracavity power, respectively, Eq. (5) reduces to the well-known steady-state cubic equation of dispersive optical bistability [115, 116]:

$$X = Y^3 - 2\Delta Y^2 + (\Delta^2 + 1)Y. \tag{7}$$

In Fig. 3(a), we plot the curves $Y(X)$ for different values of $\Delta$. One can see that $Y(X)$ is single valued for $\Delta \leq \sqrt{3}$, whereas for $\Delta \geq \sqrt{3}$, it has an S-shape, similarly to a bistable hysteresis cycle. Figure 3(b) presents the curves $Y(\Delta)$ for different values of $X$. One can see that the cavity resonance is displaced and that the cavity response becomes multi-valued when $X$ becomes sufficiently large, which corresponds to the nonlinear Kerr-tilt of the cavity resonance towards positive detuning for a constant pump power.

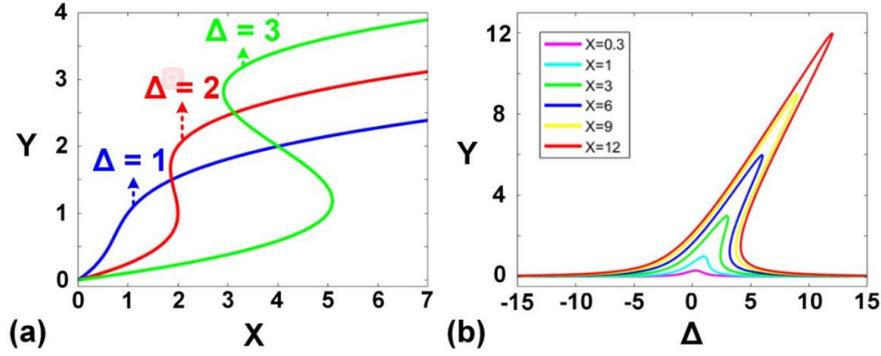

Fig. 3. Micro-comb dynamics simulated by LLE. (a) Intracavity power $Y$ as a function of the normalized pump power $X$ for different values of cavity detuning $\Delta$ = 1, 2, 3. (b) Intracavity power $Y$ as a function of cavity detuning $\Delta$ for different normalized pump values $X$ = 0.3, 1, 3, 6, 9, 12.

As discussed in detail elsewhere [44, 46, 51, 115], when the detuning $\Delta$ is adiabatically ramped along the tilted cavity response (mirroring the experimental process of detuning the laser frequency through the resonance from blue to red wavelengths), different types of solutions for Eq. (7) correspond to different regimes of comb dynamics, including primary combs, chaos, and cavity solitons. The analysis of the solutions of Eq. (7) and their bifurcations provides significant insights into both the comb characteristics and formation dynamics.

The primary combs, also called roll patterns, correspond to the initial state of comb formation that emerges from noise after breakdown of an unstable flat background via MI. When the power of the CW pump is above a certain threshold, the comb formation initiates, leading to the generation of primary sidebands separated from the pump by multiple free spectral ranges (FSRs). This state preferably occurs in the regime of anomalous dispersion, but it can also be sustained in the normal dispersion, although under very marginal conditions [112, 117].

Chaotic behaviour can be generated by destabilizing the primary combs, when comb becomes single-FSR spaced and the amplitudes evolve chaotically. The Kerr comb features many spurious modes that are subjected to fluctuations over timescales of the order of a photon lifetime. Chaos can also be induced by unstable solitons, where the Kerr comb is characterized by the pseudorandom emergence of very sharp and powerful peaks, which can be used for the generation of optical Rogue waves [118, 119]. However, since the pseudo-random spectra of chaotic combs are nonstationary and only partially coherent, they are not the most suitable micro-combs for applications where high coherence is required.

Cavity solitons result from balances between both nonlinearity and dispersion, and between gain and dissipation. In the spectral domain, Kerr combs corresponding to cavity solitons display up to several hundred phase-locked, highly coherent modes, and are the most widely used in practical applications. Various types of solitons such as bright [30, 120], dark [93, 121], as well as breathers [106, 122] have also been investigated in micro-combs. It should also be noted that solitons are typically not influenced by boundary conditions as long as their pulse widths are substantially shorter than the cavity lengths [57, 107]. Therefore, it is possible to simultaneously excite several solitons inside the same resonator, and they can remain uncoupled given that their pulse widths remain much shorter than the separation between them.

## III. RF PHOTONICS BASED ON MICRO-COMBS

RF photonics, which deals with applications in the RF to microwave and MMW range using photonic technologies, has found wide applications in wide-bandwidth analogue signal transmission, control of phased arrays, analogue-to-digital conversion, spectral filtering, and RF signal processing [1-3, 123-125]. RF photonic techniques are typically implemented through the modulation of optical carriers with the RF signals to be processed, then manipulating the modulated optical signals with low-loss photonic approaches and converting the processed optical signals back to the RF regime [3, 126]. Owing to compact device footprints and large number of available wavelength channels, micro-combs offer powerful new ways to achieve RF signal generation, filtering, and processing with reduced cost and improved performance.

### 1. RF photonic oscillators

Microwave and MMW sources are key elements for wireless communications systems, metrology, radar, radio astronomy, satellites, GPS navigation, spectroscopy [24, 127]. In practical applications, RF sources with low phase noise and high spectral purity are highly desirable. Electronic oscillators draw their spectral purity from the high Q factors of the resonators in their circuits, which significantly degrade at 10 to 100 GHz and beyond [38]. In contrast, RF photonic oscillators can be used for generating microwave and MMW signals with frequencies limited only by the bandwidth of the photodetector (PD), typically achieving the highest spectral purity at a microwave frequency [22, 128]. A state-of-the-art approach to RF signal generation is based on optical frequency division, where the beat note of a stabilized OFC interacting with a mode-locked laser is detected, generating a spectrally-pure RF signal [22, 128, 129]. A key drawback of these RF photonic oscillators is that they are generally bulky and complex, which significantly limits their use outside the laboratory [38, 128], while micro-combs hold great promise for transforming these lab-scale oscillators to the chip-scale level.

The high potential of micro-combs in implementing compact RF photonic oscillators has been recognized even since their early demonstration [79, 130]. By using a fast photodiode to detect the beat note of a phase-locked micro-comb, stable and spectrally pure RF signals can be obtained. Early experiments showed that the phase noise of the RF signal generated by a free-running micro-comb is comparable to an open-loop, high-performance electronic RF oscillator [37]. Subsequently, detailed studies on the phase noise of micro-combs were conducted by A. Matsko and L. Maleki, et al [38, 131-134]. They found that the phase noise of the micro-comb beat note is affected by several parameters, including the intensity noise of the pump laser, the Q-factor of the micro-resonator, thermal fluctuations, pump-resonance detuning, shot noise, and the comb mode-locking mechanism.

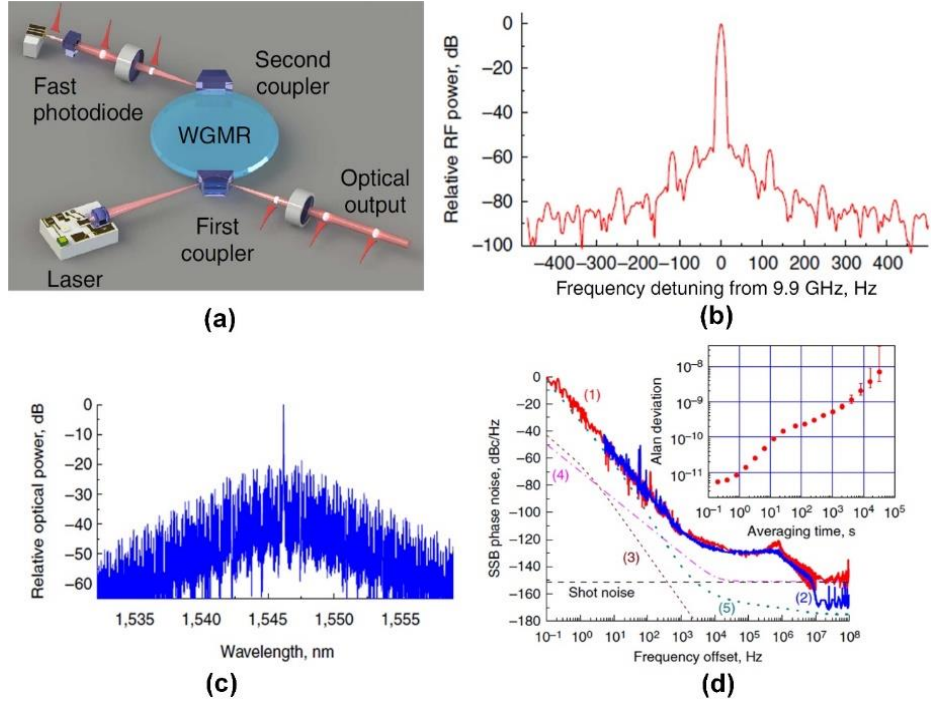

Fig. 4. RF photonic oscillator based on an $MgF_2$ micro-comb. (a) Schematic diagram of the setup of the RF photonic oscillator based on optical hyper-parametric oscillation in an $MgF_2$ crystalline WGM resonator. (b) Spectrum of the generated microwave signal measured with 1-kHz span and 9-Hz resolution bandwidth. (c) Spectrum of the micro-comb for generating the microwave signal. (d) Phase noise of the 1-GHz RF signal generated by the micro-comb without (red line, (1)) and with (blue line, (2)) a narrow-band RF filter placed after the PD. Curves (3), (4), and (5) theoretically describe thermo-refractive noise, quantum noise, and sensitivity of the phase noise measurement system, respectively. Inset shows the Allan deviation of the RF signal. Adapted from Ref. [38].

Figure 4 shows experimental results of an RF photonic oscillator based on a broadband phase-locked micro-comb from an $MgF_2$ WGM resonator with an intrinsic Q factor of $\sim 5 \times 10^9$ [38]. Figures 4(a) and (b) depict a schematic of the setup, along with the RF spectrum of the generated 9.9-GHz microwave signal. The optical spectrum of the micro-comb used to generate the microwave signal is shown in Fig. 4(c). The pump laser was a semiconductor DFB laser, with the frequency locked to a selected resonator mode through self-injection locking. The temperature of the resonator was stabilized to milli-Kelvin (mK) levels. A drop port was used to retrieve the comb for microwave generation [38]. Due to the resonator's filtering property, the noisy steady-state background from the pump laser can be suppressed by using the drop port. The measured single sideband power spectral density of the microwave phase noise is −60 dBc/Hz at 10 Hz, −90 dBc/Hz at 100 Hz, and −170 dBc/Hz at 10 MHz, as shown in Fig. 4(d). The Allan deviation of the signal is $10^{-11}$ at an integration time of 1s. Such a performance is superior to chip-scale laser-based RF photonic oscillators. The phase noise at small offset frequencies below 1 kHz is limited by fluctuations of the resonator frequency. The noise floor above 10 MHz is limited by shot noise, which can be further reduced by inserting a narrow-band RF filter after the PD. For intermediate frequencies from 1 to 10 MHz, the phase noise is mainly induced by a transfer of the laser relative intensity noise onto the microwave phase modulation through comb dynamics. The theoretical limits resulting from quantum vacuum fluctuations and thermodynamic fluctuations of the resonator temperature and volume are also shown in Fig. 4(d). By employing a better thermal and mechanical stabilization of the system and reducing the laser relative intensity noise, the spectral purity could be further improved.

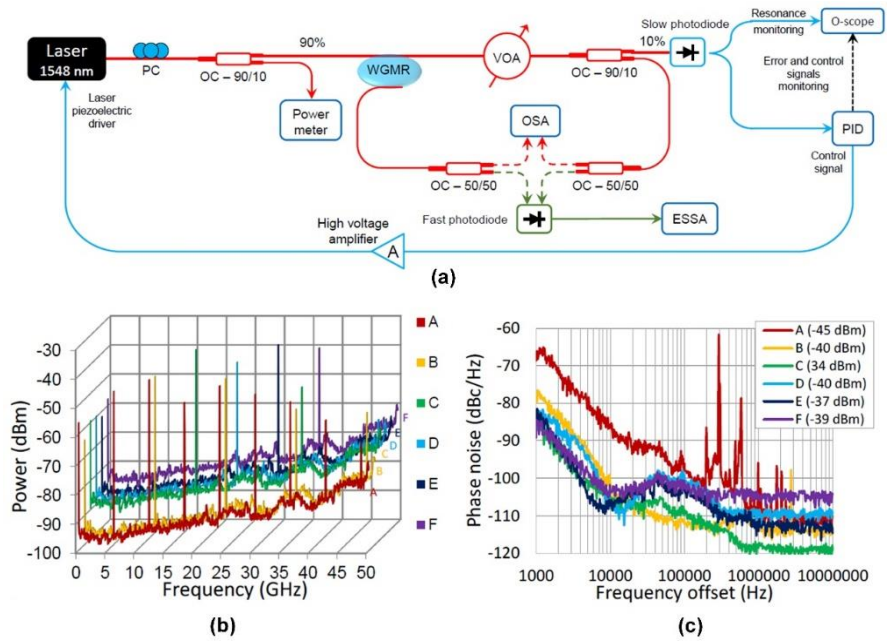

Fig. 5. Multi-frequency RF photonic oscillator based on an MgF$_2$ micro-comb. (a) Experimental setup including the side-of-fringe laser locking loop used to probe the different Kerr comb patterns generated in the MgF$_2$ resonator. In red: optical path; In green: RF path; In blue: low frequency path for the locking loop. WGMR: WGM resonator. PC: polarization controller. VOA: variable optical attenuator. OC: optical fibre coupler. O-scope: oscilloscope. PID: proportional-integral-derivative controller. OSA: optical spectrum analyser. ESSA: electrical signal and spectrum analyser. (b) Spectra of the RF signals generated by different combs corresponding to different optical modes of the MgF$_2$ WGMR. A: 6 GHz, B: 12 GHz, C: 18 GHz, D: 24 GHz, E: 30 GHz, F: 36 GHz. (c) Phase noise spectra of the main beat notes of the different RF signals in (b). Adapted from Ref. [135].

In another report of an RF photonic oscillator based on micro-combs [135], an MgF$_2$ WGM disk-resonator with a 12-mm diameter was used to generate RF signals at a range of frequencies. The intrinsic and loaded Q factors were $6\times10^8$ and $1.6\times10^8$, respectively. The experimental setup is shown in Fig. 5(a). The blue slope of the identified optical mode was probed by using the side-of-fringe laser locking technique, where the partially-resonant laser frequency changes were translated into intensity changes and then fed back to the laser to stabilize it on the required mode slope. Figure 5(b) shows the spectra of the RF signals generated by the Kerr combs with different comb spacings of ~6 GHz (A), ~12 GHz (B), ~18 GHz (C), ~24 GHz (D), ~30 GHz (E), and ~36 GHz (F), respectively. These various patterns of Kerr combs were generated and identified while the laser was gradually tuned through the resonance in the identified optical mode of the MgF$_2$ WGM resonator. The translation from one pattern to another depended on the laser-mode locking state. Both combs used to generate RF signals A and B were not pure primary combs, whereas for the combs used to generate RF signals C, D, E, and F, the generated combs were genuinely primary. The measured phase noise spectra of the RF signals in Fig. 5(b) are given in Fig. 5(c), along with the obtained RF power for the generated signals. These phase noise results confirm the better performance for purely primary combs. For the generated RF signals with frequencies over 10 GHz (C, D, E, F), phase noise levels below −100 dBc/Hz at a frequency offset of 10 kHz with respect the different carriers were obtained.

## 2. RF photonic true time delay lines

In modern radar and communications systems, RF photonic true time delay lines (TTDLs), which introduce multiple progressive time delays by means of RF photonic technologies, are basic building blocks with wide applications in phased array antennas (PAAs), RF photonic filters, analog-to-digital or digital-to-analog conversion, and arbitrary waveform generation [136, 137]. For RF photonic TTDLs, increasing the number of delay channels can lead to a significantly improved performance. For instance, in PAAs, the number of radiating elements determines the beamwidth, and so an improved angular resolution can be achieved by increasing the channel number [138]. Consequently, RF photonic TTDLs with a large number of channels are highly desirable for PAAs with high angular resolution. In conventional methods, discrete lasers arrays [139, 140] or FBG arrays [138] are employed towards implementing multiple OTTD channels, where system cost and complexity are significantly increased with the rising channel number, thus greatly limiting the number of available channels in practical systems. In contrast, for a broadband Kerr optical micro-comb generated by a single micro-resonator, a large number of high-quality wavelength channels can be simultaneously provided for the RF photonic TTDL, thus greatly reducing the size, cost, and complexity of the system.

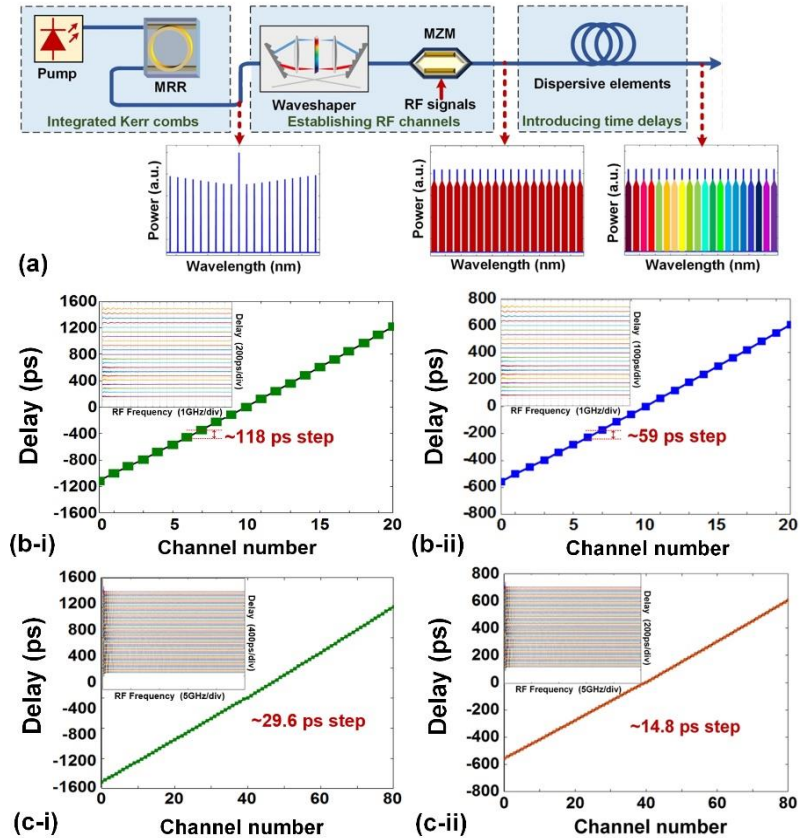

Fig. 6. RF photonic TTDLs based on Hydex micro-comb sources. (a) Schematic diagram. MRR: micro-ring resonator. MZM: Mach-Zehnder modulator. (b) Measured time delay responses of a 21-channel TTDL with (i) 4-km-long dispersive SMF and (ii) 2-km-long dispersive SMF. (c) Measured time delay responses of an 81-channel TTDL with (i) 4-km-long dispersive SMF and (ii) 2-km-long dispersive SMF. Inset in (b) and (c) show the flat delays over a wide RF frequency range.

Figure 6(a) shows a schematic diagram of a multi-channel RF photonic TTDL with a micro-comb source. The OTTD is constructed with three modules. The first module generates the micro-comb with a large number of wavelength channels using a micro-resonator. The second module shapes the micro-comb and directs it to a Mach-

Zehnder modulator (MZM), where replicas of the input RF signal are generated on each wavelength to establish multiple RF channels. After that, time delays were introduced between those RF channels by dispersive elements in the third module. The measured time delay responses of a 21-channel RF photonic TTDL based on the micro-comb generated by a Hydex MRR with a FSR of ~200 GHz are shown in Fig. 6(b). We adopted two different lengths of the dispersive single mode fibre (SMF), and got two different delay steps of ~118 ps/channel and ~59 ps/channel. As shown in Fig. 6(c), we also measured the time delay responses of an 81-channel RF photonic TTDL based on the micro-comb generated by another Hydex MRR with a FSR of ~49 GHz and got two different delay steps of ~29.6 ps/channel and ~14.8 ps/channel. The time delay steps were jointly determined by the frequency spacing of the generated comb and the accumulated dispersion of the single mode fibre (SMF), so that we got four different delay steps in Figs. 6(b) and (c).

One of the important applications of an RF photonic TTDL is in phased array antenna (PAA) systems, where optical signals on selected channels are separately converted into the electrical domain and fed to an antenna array to form radiating elements. Considering that the antenna array is a uniformly spaced linear array with an element spacing of $d_{PAA}$, the steering angle $\theta_0$ of the PAA can be given as [141]

$$\theta_0 = \sin^{-1} \frac{c \cdot \tau}{d_{PAA}}, \tag{8}$$

where $c$ is the speed of light in vacuum, and $\tau = mT$ ($m = 1, 2, 3, …$) is the time delay difference between adjacent radiating elements, with $T$ denoting the time delay difference between adjacent channels. The steering angle can be tuned by adjusting $\tau$, i.e., by changing the length of the dispersive medium or simply selecting every $m_{th}$ channel as radiating elements. The corresponding array factor (AF) of the PAA can be expressed as [141]

$$AF(\theta, \lambda) = \frac{\sin^2[M\pi(d_{PAA}/\lambda)(\sin\theta - c \cdot mT/d_{PAA})]}{M^2 \sin^2[\pi(d_{PAA}/\lambda)(\sin\theta - c \cdot mT/d_{PAA})]} \tag{9}$$

where $\theta$ is the radiation angle, $M$ is the number of radiating elements, and $\lambda$ is the wavelength of the RF signals. The angular resolution of the PAA is the minimum angular separation at which two equal targets at the same range can be separated. It is determined by the 3-dB beamwidth of the PAA, which can be approximated [142] as $\theta_{3dB} = 102 / M$, thus indicating that the angular resolution would greatly increase with a larger number of radiating elements, and this can be realised via micro-combs that can provide numerous wavelength channels for beam steering.

Figure 7(a) depicts the relationship between the number of radiating elements ($M$) and the 3-dB beamwidth ($\theta_{3dB}$) of the PAA, where smaller beamwidths can be obtained when $M$ increases, therefore suggesting, given the large number of available channels provided by the micro-comb based RF photonic TTDL, that the angular resolution for the PAA can be greatly enhanced. The improved angular resolution for increased $M$ can also be observed in Fig. 7(b), which shows the AFs of the PAA for different $M$ calculated based on Eq. (9). To achieve a tunable beam steering angle, we selected every $m_{th}$ ($m = 1, 2, 3, …$) wavelength of the TTDL by using the waveshaper, in such a way that the time delay ($\tau$) between the radiating elements could be varied with a step size of $T$. Figs. 7(c) and (c) show the calculated AFs for $m$ ($m$ = Channel number / $M$) varying from 1 to 7 based on a 200-GHz-FSR Hydex micro-comb and varying from 1 to 27 based on a 49-GHz-FSR Hydex micro-comb, respectively. As can be seen, on one hand, the 49-GHz-FSR Kerr comb enables a much larger $M$ as $m$ varies, thus leading to a much smaller beamwidth and to a greatly enhanced angular resolution. On another hand, finer tuning steps (from 1.0° to 14.7°) as well as a large tuning range (142.7°) of the beam steering angle are available due to the increased number of *m*. Moreover, PAAs based on such RF photonic TTDLs can also achieve wide instantaneous bandwidths without beam squints (the variation of beam steering angle as a function of RF frequency). As indicated in Figs. 7(e) and (f), the beam steering angle remains the same while the RF frequency varies. These experimental results confirm that PAAs based on RF photonic TTDLs with micro-comb sources feature greatly improved angular resolution, a wide tunable range in terms of beam steering angle, and a large instantaneous bandwidth.

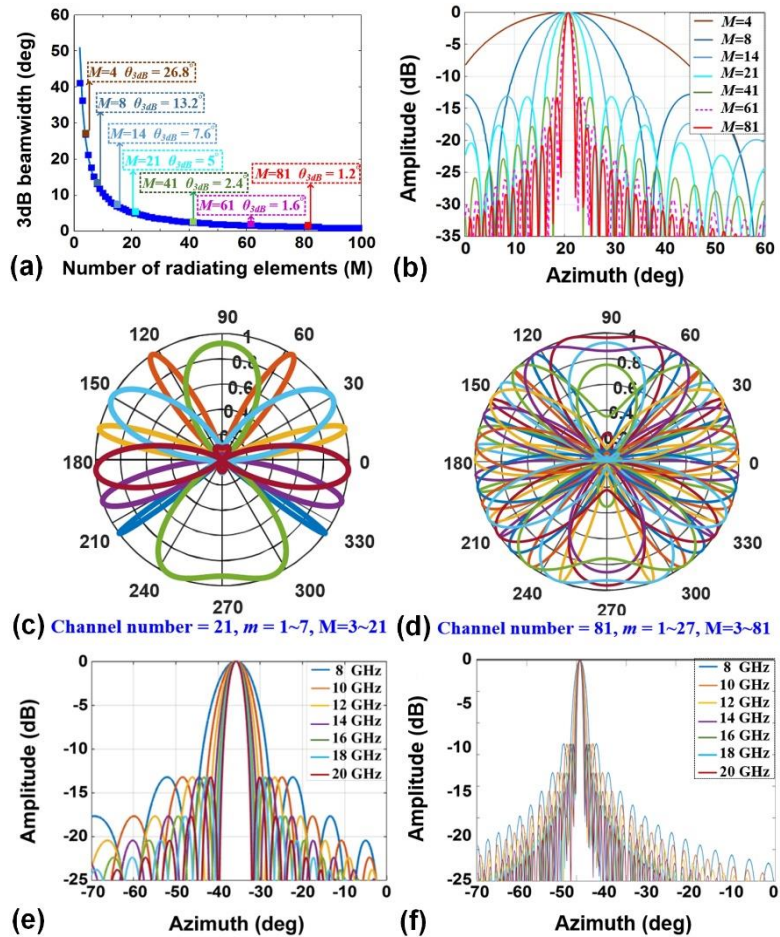

Fig. 7. (a) Relationship between the number of radiating elements (*M*) and the 3dB beamwidth of the PAA. (b) Calculated AFs of the PAA with *M* varying from 4 to 81 based on a 49-GHz-FSR Hydex micro-comb. (c) Calculated AFs of the PAA with *m* varying from 1 to 7 based on a 200-GHz-FSR Hydex micro-comb. (d) Calculated AFs of the PAA with *m* varying from 1 to 27 based on a 49-GHz-FSR Hydex micro-comb. (e) Calculated AFs of the PAA with various RF frequencies based on a 200-GHz-FSR Hydex micro-comb. (f) Calculated AFs of the PAA with various RF frequencies based on a 49-GHz-FSR Hydex micro-comb.

3. RF photonic filters

Another key application of RF photonics is RF photonic filters – photonic subsystems that perform equivalent functions to those of ordinary RF filters in RF systems. RF Photonic filters offer several competitive advantages over their electrical counterparts, including low loss, large filter bandwidths, reconfigurable filter shapes, fast tunability, and strong immunity to electromagnetic interference (EMI) [143]. Among the various schemes to implement RF photonic filters, tapped delay-line filters, also called transversal filters, have attracted great interest due to their high reconfigurability in terms of filtering shapes [23, 24, 144]. In a photonic delay-line filter, discrete frequency samples of the optical signal containing the RF modulation are time delayed, weight tailored, summed together, and at the end detected by a PD to generate the RF output. This is essentially equivalent to the digital filters in signal processing (DSP), but implemented by photonic hardware. The transfer function can be expressed as [144]:

$$H(\omega) = \sum_{n=0}^{N-1} a_n e^{-j\omega nT},  \qquad (10)$$

where $\omega$ is the angular frequency of the input RF signal, $N$ is the number of taps, $a_n$ is the tap coefficient of the $n_{th}$ tap, and $T$ is the time delay between adjacent channels. By employing calculated tap coefficients $a_n$ on each wavelength channel, reconfigurable RF photonic filters with arbitrary spectral transfer functions can be implemented. Since the photonic delay-line filters utilize optical dispersive delay lines at different wavelength channels as distinctive taps of the delay-line filters, they require a large number of wavelength channels. Further increasing that number can lead to improved performance of RF photonic delay-line filters such as better filtering Q factors and time-bandwidth products. OFCs have demonstrated a great potential to meet this requirement [23, 24]. High-quality, broadband multi-wavelength sources for RF photonic filtering have typically been based on mode-locked fibre lasers [145] and OFCs achieved by electro-optical (EO) modulation [23, 146]. Recent demonstrations [147] have shown that micro-combs hold particularly significant potential for high-performance transversal filtering, offering greatly reduced size, cost, and complexity.

Xue et al. reported the first demonstration of RF photonic filters based on a micro-comb [147]. The experimental setup is shown in Figure 8(a). The Kerr comb generated by a $Si_3N_4$ MRR in the Kerr comb generation module was tailored by pulse shaper 1 before being launched into the photonic RF filter module. In the photonic RF filter section, the tailored comb was split into two branches. One branch was modulated by an RF input signal and then tailored by pulse shaper 2. The other branch was passed without change except for a small tunable delay. Due to the interference between these two branches, one can selectively suppress unwanted RF passbands by programming pulse shaper 2. The optical spectrum of the micro-comb generated by the $Si_3N_4$ MRR is shown in Fig. 8(b). The comb spacing is ~231.3 GHz, which makes it possible to combine optical and RF filtering to suppress unwanted RF passbands including the image and periodic passbands. Figure 8(c) shows the measured single-bandpass RF photonic filter when the centre frequency was tuned to 2.5 GHz, 7.5 GHz, 12.5 GHz, and 17.5 GHz, respectively. The simulation results obtained by assuming that the unwanted passbands are not suppressed are also shown. The experimental and simulated results agree well with each other in the designed bandpass regions, and all the unwanted passbands are effectively suppressed in the experimental results. Figure 8(d) shows the flat-top single-passband RF photonic filters with tunable centre frequency from 2.5 GHz to 17.5 GHz implemented by the same micro-comb. The full width at half maximum (FWHM) of the RF passband is about 4.3 GHz.

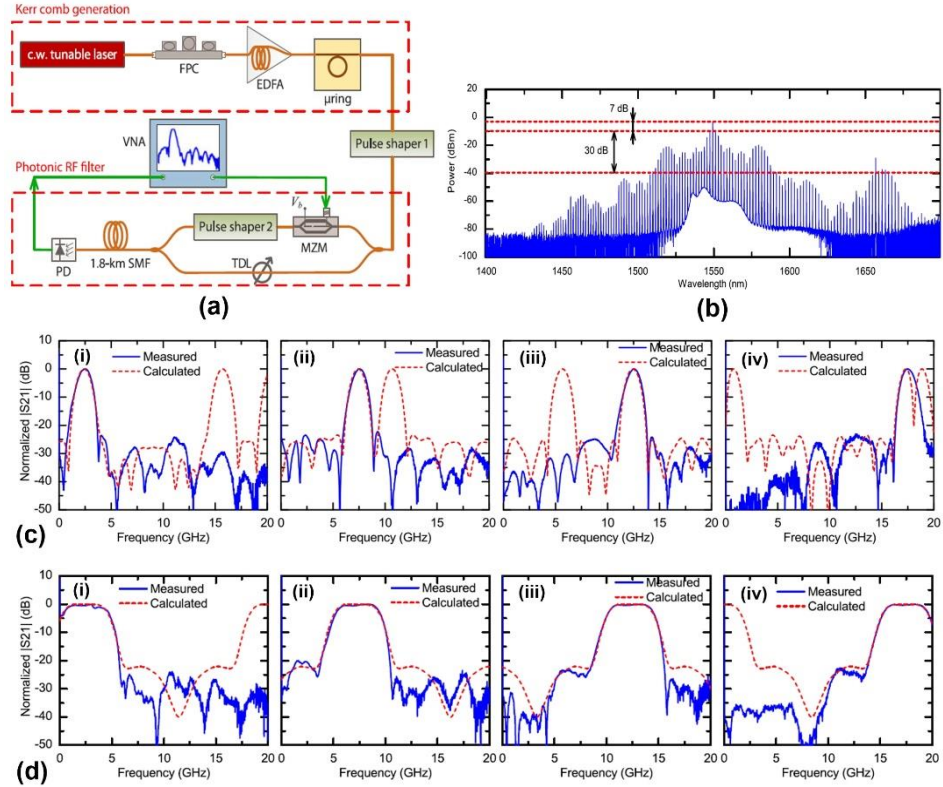

Fig. 8. Single bandpass RF photonic filter based on a $Si_3N_4$ micro-comb. (a) Experimental setup. FPC: fibre polarization controller. EDFA: erbium-doped fibre amplifier. MZM: Mach-Zehnder modulator. TDL: tunable delay line. SMF: single-mode fibre. PD: photodetector. (b) Spectrum of the micro-comb generated by a $Si_3N_4$ MRR. (c) RF transfer function when the passband is the Fourier transform of a Hamming window and the centre frequency is tuned to (i) 2.5 GHz, (ii) 7.5 GHz, (iii) 12.5 GHz, and (iv) 17.5 GHz. (d) RF filtering responses when the passband is flat-top and the centre frequency is tuned to (i) 2.5 GHz, (ii) 7.5 GHz, (iii) 12.5 GHz, and (iv) 17.5 GHz. Adapted from Ref. [147].

Figure 9(a) shows a schematic diagram of a reconfigurable RF photonic filter based on a micro-comb generated by a Hydex MRR. The generated micro-comb was amplified and shaped by a waveshaper to get weighted taps according to calculated tap coefficients. To increase the accuracy of comb shaping, a real-time feedback control path was adopted to read and shape the comb lines' power accurately. A 2×2 balanced MZM simultaneously modulated the input RF signal on both positive and negative slopes to achieve tap coefficients with opposite algebraic signs. Figure 9(b) shows the measured RF amplitude response of an "all-ones" RF filter with all tap coefficients set to unity. One can see that the Q factor increases as the tap number passes from 4 to 14, demonstrating that it can be greatly improved with the large tap number provided by the micro-comb. The measured RF amplitude response of a tunable RF bandpass filter with different centre frequencies is shown in Fig. 9(c). Tunable centre frequencies from ~25.27% to ~57.98% of the Nyquist frequency were achieved by programming the weights of taps based on the Remez algorithm [148]. RF photonic filters based on micro-combs can realise versatile filter shapes by programming the tap weights according to calculated tap coefficients. Figures. 9(d)−(h) show five typical filter shapes achieved by the same RF photonic filter based on the Hydex micro-comb, including a half-band highpass filter, a half-band lowpass filter, a band-stop filter, a Nyquist filter, and a raised cosine filter. It can be seen that all five filters feature the responses expected from simulations, which confirms the high performance and reconfigurability of micro-comb based RF photonic filters.

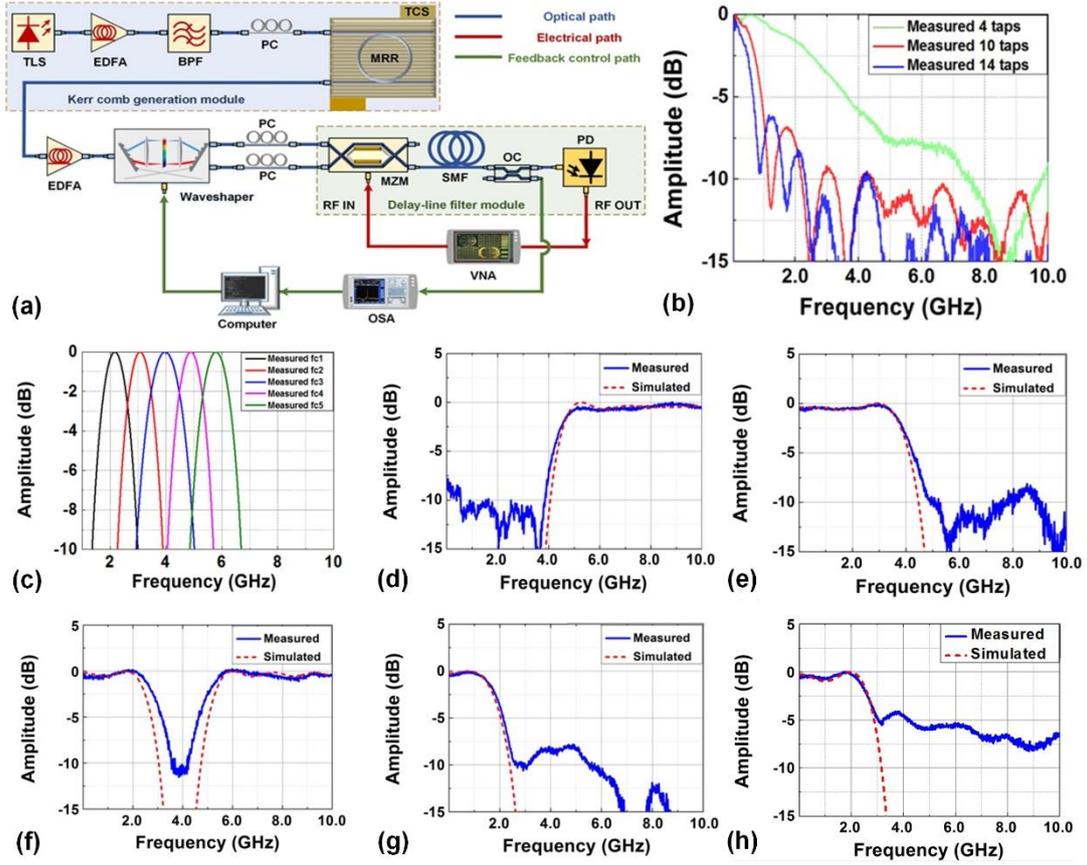

Fig. 9. Reconfigurable RF photonic filter based on a Hydex micro-comb. (a) Schematic illustration. TLS: tunable laser source. EDFA: erbium-doped fibre amplifier. PC: polarization controller. BPF: optical bandpass filter. TCS: temperature controller stage. MZM: Mach-Zehnder modulator. SMF: single mode fibre. OC: optical coupler. PD: photodetector. OSA: optical spectrum analyser. VNA: vector network analyser. (b) Measured RF amplitude responses of the all-ones filter with different number of taps. (c) measured RF amplitude responses of a bandpass filter with tunable centre frequencies. (d)−(h) Measured and simulated RF amplitude responses of half-band highpass, half-band lowpass, band-stop, Nyquist, and raised-cosine filters, respectively.

## 4. RF Hilbert transformer

In section 3.3, we discussed RF photonic delay-line filters based on micro-combs. In the temporal domain, these delay-line filters can be properly designed to achieve ultrahigh-speed RF photonic signal processing. In comparison with electronic counterparts, RF photonic signal processing technologies can overcome the bandwidth bottleneck and perform functions that are very complex or even impossible to realise in the RF domain. Moreover, RF photonic signal processing also exhibits other attractive properties, including low loss, high flexibility, and strong immunity to EMI, which are crucial for microwave and MMW processing systems [149, 150].

The distinctive feature of micro-combs for RF photonic signal processing is that they can exhibit large comb spacings, which are challenging to achieve for mode-locked lasers and OFCs using EO modulation. For RF photonic processors based on micro-combs, the operation bandwidth is theoretically limited by half of the comb spacing, or the Nyquist zone [144]. Therefore, the large spacing of micro-combs allows a large Nyquist zone, which can be as large as 100's of GHz − well beyond the processing bandwidth of electronic devices. The high reconfigurability of RF delay-line filters also allows versatile processing functions and operation bandwidths. By simply programming and shaping the comb lines according to specific tap coefficients, the same setup can be employed to achieve multiple processing functions. Note that such a high degree of reconfigurability cannot be typically obtained by passive silicon counterparts [87, 151, 152], thus making these multi-functional RF photonic processors more suitable for the diverse computing requirements in practical applications. Moreover, since the generated comb only serves as a multi-wavelength source for the subsequent transversal filter, in which the optical power from different taps is detected incoherently by the photo-detector, achieving rigorous comb coherence is not crucial and so these RF photonic processors are able to work under relatively incoherent conditions.

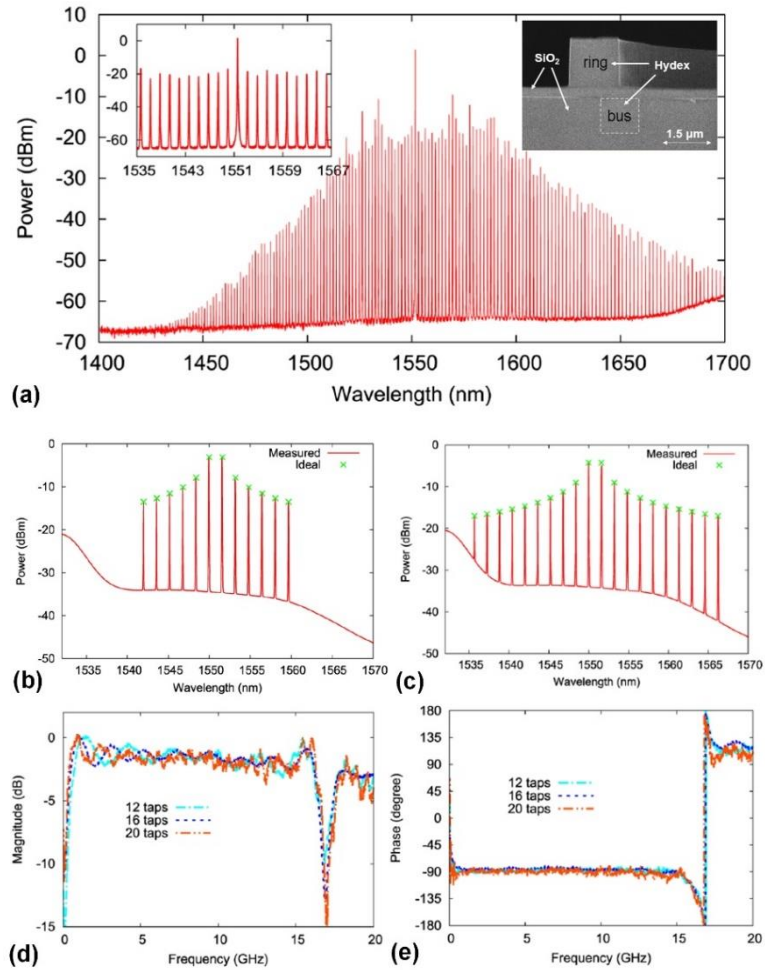

Fig. 10. An RF photonic Hilbert transformer based on a Hydex micro-comb. (a) Optical spectrum of the generated micro-comb from 1400 nm to 1700 nm. Inset shows a zoom-in spectrum with a span of ~32 nm and the Hydex MRR employed for generating the comb. (b)−(c) Measured optical spectra (red solid) of the shaped optical combs and ideal tap weights (green crossing) for the filter with 12 and 20 taps, respectively. (d)−(e) Measured RF amplitude and phase responses of the RF photonic Hilbert transformer, respectively. Adapted from Ref.[39].

An RF photonic Hilbert transformer based on a Hydex micro-comb was demonstrated in 2015 [39], with up to 20 taps and more than a 5-octave 3-dB bandwidth. In previous demonstrations of Hilbert transformers based on photonic transversal filtering, filter taps were achieved by employing an array of discrete CW laser sources, which limited the number of available taps to only four and resulted in less than a 3-octave bandwidth. On the other hand, for the Hilbert transformer based on a micro-comb, the wide spectral width and large frequency spacing of micro-combs allows for a large number of high-quality filter taps, thus leading to much broader RF bandwidths than those typically obtained with standard RF circuits. The spectrum of a Kerr optical comb generated by a Hydex MRR with a FSR of ~200 GHz is shown in Fig. 10(a). The shaped optical spectra with 12 and 20 taps are presented in Figs. 10(b) and (c), respectively. Figure 10(d) shows the measured RF amplitude frequency response for the Hilbert transformer with 12, 16 and 20 taps. All three filters have the same null frequency at ~16.9 GHz, corresponding to a tap spacing of ~59 ps determined by the delay difference between adjacent comb lines. An increased number of filter taps leads to a larger operation bandwidth. The Hilbert transformer with 20 taps had a 3-dB bandwidth extending from ~0.3 GHz to ~16.4 GHz, corresponding to more than 5 octaves, which can be further increased by using more comb lines. Fig. 10(e) presents the measured phase response of the Hilbert transformer with different numbers of taps, showing similar responses. Each shows a relatively constant phase of near −90° within the pass-band.

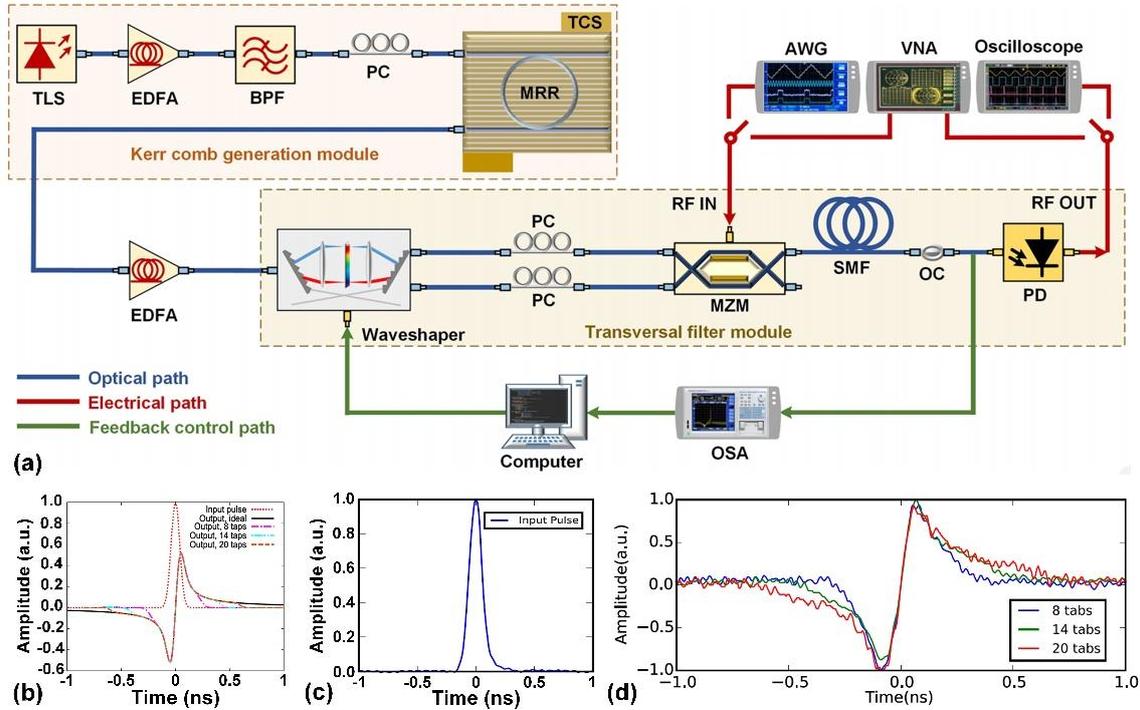

Fig. 11. Real-time Hilbert transformation based on a Hydex micro-comb. (a) Schematic illustration of the experimental setup. TLS: tunable laser source. EDFA: erbium-doped fibre amplifier. PC: polarization controller. BPF: optical bandpass filter. TCS: temperature control stage. MZM: Mach-Zehnder modulator. SMF: single mode fibre. OC: optical coupler. PD: photo-detector. OSA: optical spectrum analyser. VNA: vector network analyser. AWG: arbitrary waveform generator. (b) Simulated input and output waveforms of the temporal Hilbert transformer with 8, 14, and 20 taps. (c) Measured temporal waveform of a base-band Gaussian-like input pulse. (d) Measured output waveforms of the input signal in (c) after being processed by the Hilbert transformer with 8, 14, and 20 taps.

To further investigate the temporal responses of the Hilbert transformer based on the Hydex micro-comb, we used the experimental setup in Fig. 11(a) to perform real-time Hilbert transformation for a base-band Gaussian input signal. Figures 11(b) shows the simulated input and output waveforms of the temporal Hilbert transformer with 8, 14, and 20 taps. One can see that the deviation between the output waveform and the ideal Hilbert transformation is inversely proportional to the number of taps, as reasonably expected. The measured temporal waveform of a base-band Gaussian-like input signal produced by an arbitrary waveform generator (AWG) is presented in Fig. 11(c), where the output waveforms measured after being processed by the Hilbert transformer are shown in Fig. 11(d). One can see that such waveforms are consistent with their theoretical counterparts, which confirms effective temporal Hilbert transformation using the micro-comb based RF photonic processors.

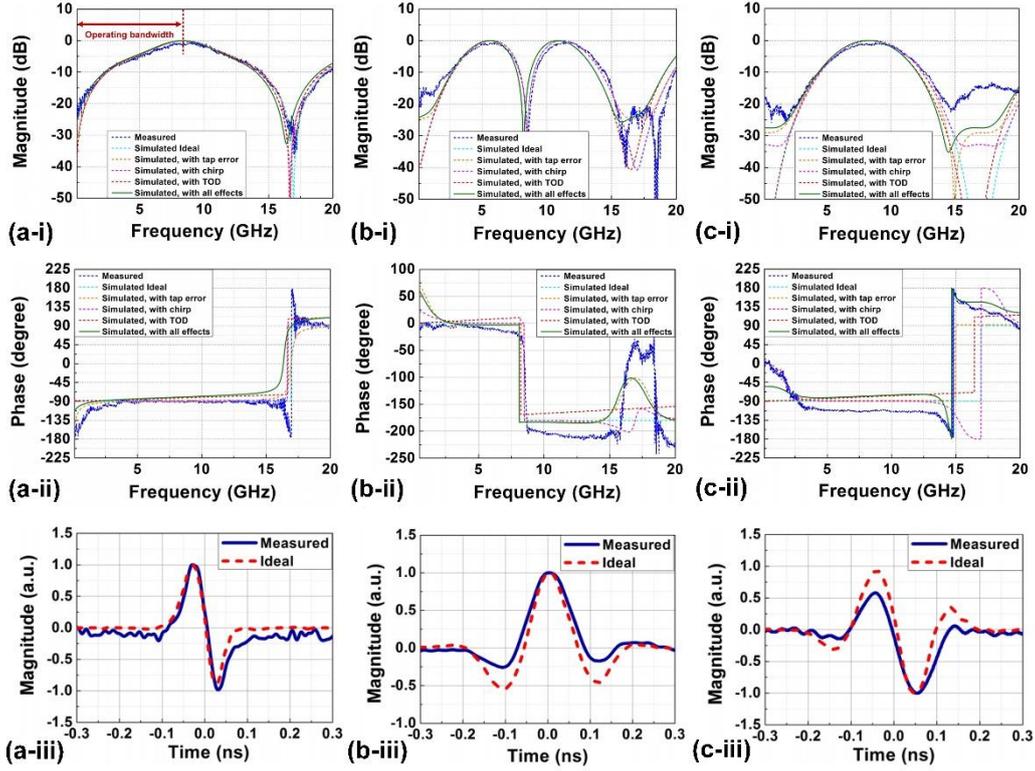

Fig. 12. Reconfigurable RF photonic intensity differentiator based on a Hydex micro-comb. (a) Measured and theoretical (i) RF amplitude response, (ii) RF phase response, and (iii) temporal output waveform after first-order intensity differentiation. (b) Measured and theoretical (i) RF amplitude response, (ii) RF phase response, and (iii) temporal output waveform after second-order intensity differentiation. (c) Measured and theoretical (i) RF amplitude response, (ii) RF phase response, and (iii) temporal output waveform after third-order intensity differentiation. Adapted from Ref.[153].

## 5. RF photonic intensity differentiators

Xu et.al [153] recently reported a reconfigurable RF photonic intensity differentiator based on a Hydex micro-comb. The differentiator consists of two main blocks: the micro-comb generation module based on a Hydex MRR and a transversal filter module for reconfigurable intensity differentiation. By programming and shaping the power of individual comb lines according to a set of tap weights calculated by the Remez algorithm, first, second, and third-order, as well as square root intensity differentiation was achieved. Figures 12(a-i), (b-i), and (c-i) show the measured and simulated amplitude responses associated to the respective intensity differentiators. The corresponding RF phase responses are shown in Figs. 12(a-ii), (b-ii), and (c-ii). It can be seen that all the experimental results exhibit the responses expected from ideal differentiation. The deviations can be attributed to the tap weight error during the comb shaping, the chirp induced by the MZM, and the third-order dispersion (TOD) of the fibre. The simulated and measured temporal output waveforms after the first, second, and third-order intensity differentiation for a base-band Gaussian-like input signal are shown in Figs. 12(a-iii), (b-iii), and (c-iii), respectively. One can see that the measured waveforms closely match their theoretical counterparts, indicating good agreement between experiment and theory. Unlike the field differentiators that yield the derivative of a complex optical field [3, 151], waveforms showing differentiations of the intensity profile of the input RF signal were observed, indicating that intensity differentiation was indeed successfully achieved. For first-order, second-order, and third-order differentiation, the calculated root mean square errors (RMSEs) between the measured and the theoretical curves are ~4.15%, ~6.38%, and ~7.24%, respectively.

The micro-comb based transversal signal processor can also be used for square root intensity differentiation, for which the spectral transfer function can be expressed as:

$$H(\omega) \propto (j\omega)^{0.5}, \quad (11)$$

where $j = \sqrt{-1}$, and $\omega$ is the angular frequency. Figures 13(a) and (b) show the RF amplitude and phase responses of the square root intensity differentiator based on a Hydex micro-comb in our experiment. The tap coefficients corresponding to square root differentiation were calculated to be [0.0578, -0.1567, 0.3834, 1, -0.8288, 0.0985, -0.0892]. The dispersion of the ~2-km-long SMF used as a dispersive medium was ~17.4 ps/(nm·km), which corresponded to a time delay of ~59 ps between adjacent taps and yielded an effective FSR of ~16.9 GHz in the RF response spectra. Note that the device is designed for processing base-band RF signals, so the theoretical operating frequency range starts at DC and ends at half of the spectral range between DC and the notch centred at ~16.9 GHz. The measured temporal waveform of a base-band Gaussian-like input signal is shown in Fig. 13(c), with the measured and simulated output waveforms after square root intensity differentiation presented in Fig. 13(d). It can be seen that the measured output waveform agrees well with theory, with a RMSE of ~4.02%.

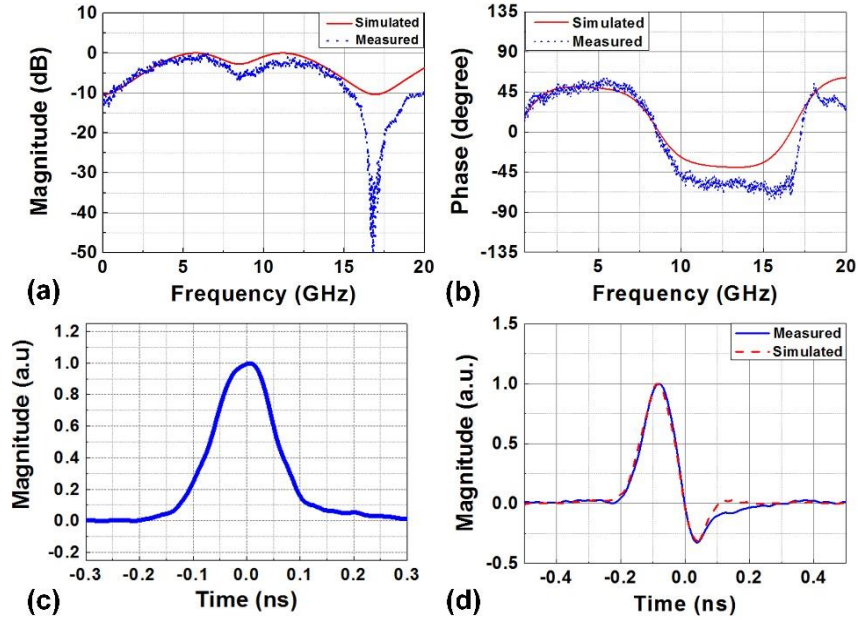

Fig. 13. RF photonic square root intensity differentiator based on a Hydex micro-comb. Measured and theoretical RF (a) amplitude and (b) phase responses of the square root intensity differentiator. (c) Measured temporal waveform of a Gaussian-like input pulse. (d) Measured and simulated output waveforms of the input signal in (c) after square root differentiation.

## IV. CHALLENGES AND LIMITATIONS

While RF photonic filters and processors based on optical micro-combs do not require a high degree of phase coherence in order to achieve high performance, this is not the case for RF photonic oscillators, where the micro-comb stability and coherence are crucial for the purity and signal-to-noise ratio (SNR) of the generated RF signals [38, 79]. Since the precise dynamic nature of micro-comb's phase and coherence typically displays a nontrivial behaviour, many challenges have been faced in achieving stable and coherent micro-combs. As discussed in Section II, cavity solitons with phase-locked modes exhibit good stability and coherence. Here, the performance and behaviour of micro-combs is strongly dependent on the excitation conditions. Therefore, it is critical to find the optimum approach to excite cavity solitons on a CW background. Another challenge is that once a soliton state is formed, the power coupled into the resonator needs to be maintained constant, otherwise the solitary state is lost. Therefore, phase locking needs to be maintained in order for the micro-combs to operate for long periods.

Continuous detuning of the pump from blue to red-shifted wavelengths has proven to be a winning strategy in searching for the excitation conditions required to achieve cavity solitons [57, 107, 154]. A significant challenge to implement such adiabatic cavity detuning for exciting cavity solitons stem from thermal nonlinear effects. The field enhancement of the micro-resonator not only lowers the energy threshold for the Kerr nonlinear effect, but also increases the thermo-optical dependence of the refractive index. Moreover, thermal responses are typically very slow [155, 156], in contrast to the ultrafast Kerr effect that underpins the mechanism for micro-comb generation [42]. This slow thermal response affects the steady state behaviour of the intracavity pump by introducing a slow dependence of the cavity detuning on pump power.

Some previous studies have demonstrated that this thermal nonlinearity can be compensated for by a thermal-locking method, which is typically achieved by adjusting the rate at which the power is increased in the cavity [36, 56, 57, 157]. This approach has been the key towards achieving stable cavity solitons in most micro-resonator materials that have positive thermal nonlinear coefficients. However, it has also been demonstrated that materials with negative thermo-optic coefficients such as $CaF_2$ could also achieve cavity soliton states [158]. The challenge for this approach has been to find the sweep rate at which the residual resonator heating can be compensated for (after the sweep is stopped) via the temperature decrease induced by the sudden loss of intra-cavity power (following the transition into cavity solitons). Another problem is that the thermal-locking state is strongly affected by the coupling state and it often translates into a decoupling if an active stabilizing loop is not in place, which would shut down the OPO. For these reasons, some other approaches have been proposed to achieve stable and coherent micro-combs, including two-colour pumping [67, 159, 160], EO modulation control [27, 161], self-injection locking [38, 64, 162], filter-driven FWM [28, 163, 164], integrated heaters [94, 165], and so forth [166, 167].

Two-colour pumping was first investigated by Strekalov et. al. in 2009 [159]. They used two optical pumps in an $MgF_2$ resonator in order to achieve better efficiency for comb generation. This approach is threshold-less since it does not rely on optical parametric generation or cavity MI, but on cascaded FWM. The generated pulses are background free, while it is still possible to find cavity solitons for such configurations. Instead of sitting on a constant background, the solitons are found to be superimposed on top of the threshold-less low-intensity pattern that is generated via FWM of the two pump modes [160]. The position of the solitons is strictly restricted by the stationary pattern, which effectively suppresses timing-jitter and interactions between different solitons [67].

Parametric EO seeding was proposed by Papp et.al. to avoid a breakup of the micro-comb into various sub-combs under certain coupling conditions [27, 161]. They experimentally investigated the properties of these kinds of micro-combs and found that the spacing of a micro-comb locks to the spacing of the EO modulated pump sidebands when the modulation frequency is close enough to the intrinsic mode spacing of the micro-resonator. EO control of the optical comb has also been exploited to achieve active stabilization of the comb lines, a property which is particularly important for high repetition rate combs that cannot be measured directly by standard electronic methods. In 2012, Diddams et.al. [168] demonstrated the measurement and stabilization of a micro-comb with a mode spacing of a 140 GHz by bridging the gap between comb modes via EO sidebands.

For prism-coupled WGM resonators such as $MgF_2$ and $CaF_2$ resonators, self-injection locking is an efficient approach to link the pump laser to a resonance of the WGM resonator [38, 64, 162]. This passive feedback mechanism based on resonant Rayleigh scattering of the WGM resonator was initially proposed by Liang et.al. to narrow the linewidth of a distributed feedback laser (DFB) with a WGM resonator [169]. A WGM resonator was used to filter a DFB laser, and some scattered light from the resonator was reflected back into the laser. The linewidth of the laser then collapsed due to self-injection locking, thus leading to light keeping within the WGM resonance. Now this approach has been widely employed for generation of extremely narrowband CW lasers as well as for the control of parametric oscillation [38, 170].

The filter-driven FWM approach proposed by Pasquazi et al in 2012 [28] consists of having a four-port MRR inserted into a laser-cavity loop to induce oscillation of the pump. Here, only a single resonance of the MRR is accurately lined up with a resonance of the outer fibre loop cavity containing the optical gain, thus experiencing amplification. Therefore, the radiation is passively linked to the resonator, and further used as a pump for the OPO in the MRR. This is an alternative approach for coherent micro-comb generation that is based on passive laser mode-locking instead of an OPO, which has proven to be a viable way to self-lock the oscillation in sub-threshold pump oscillation for quantum correlated photon state generation [40, 41, 171]. Recently, by embedding a Hydex MRR into a passive mode-locked laser structure with fibre nonlinear amplifying loop mirrors (NALMs), optical pulses in the nanosecond regime have been achieved [31]. The pulse duration was ~4.3 ns, with an overall spectral bandwidth of 104.9 MHz—more than two orders of magnitude smaller than in previous reports. The very narrow bandwidth and strong coherence make it possible to fully characterize its spectral properties in the RF domain using widely available GHz-bandwidth optoelectronic components.

Integrated heaters were recently employed by Joshi et al. [165] to assist passive thermal locking by implementing resonator rapid thermal tuning to achieve repeatable and systematic generation of low-noise single and multiple cavity soliton states. The integrated heaters allowed very fast thermal tuning of the resonances on time scales < 100 μs, which are very short compared to thermal relaxation times and so enable the accurate control of pump sweeping. Recently, an active control approach was also proposed by Yi et.al. to protect the soliton state against thermal drifting [166]. They used a complex simultaneous dynamic control of the pump laser frequency and power, within a feedback loop monitoring the average soliton output power. Using this mechanism, the soliton states were stabilized for long times [167].

Most of the current techniques for achieving mode-locking in micro-combs require sweeping the pump laser frequency. Therefore, the linewidth and amplitude noise of the pump strongly determines the quality of the micro-comb, especially in terms of phase characteristics. Tunable lasers are relatively noisy and have broader linewidths of ~100 kHz, in contrast to the fixed-wavelength lasers with record linewidths of < 40 mHz [172] that are typically based on a monolithic cavity without moving or mechanical components. These lasers can be used to significantly reduce the noise of the micro-combs in some cases where reconfigurability is not crucial.

As mentioned, RF photonic filters and processors based on micro-combs do not require a high degree of phase coherence since the combs only serve as multi-wavelength sources for the transversal filters. These RF photonic processors still perform well under relatively incoherent conditions. However, this is not to say that there are no challenges. These include the limited dynamic range and gain bandwidth of optical amplifiers, chirp and limited bandwidth of modulators, and tap errors incurred during comb shaping.

The power dynamic range of an optical amplifier is the difference between the maximum power of the generated comb lines and the power associated with the noise floor. This parameter is very important since a broader power dynamic range can lead to improved accuracy of comb shaping, and therefore will decrease deviations between performance and theory. The power dynamic range of optical amplifiers becomes even more crucial as the number of taps increases. Another important parameter of optical amplifiers that affects the micro-comb based RF photonic filters and processors is the gain bandwidth, which limits the number of available taps in practical applications. In general, optical amplifiers with high gain, bandwidth, and low noise floor are preferable in practical implementations of RF photonic filters and processors based on micro-combs.

Since EO modulators are employed to generate replicas of the input RF signals in micro-comb based RF photonic filters and processors, their parameters such as chirp and modulation bandwidth affects the filtering shape and processing performance. EO modulator chirp introduces additional tap errors before comb shaping, and so a low-chirped EO modulator is desired. Like the influence of the gain bandwidth of optical amplifiers, the modulation bandwidth of modulators also restricts the number of available taps. Actually, due to these bandwidth limitations, only a small portion of the generated comb spectrum is used in Refs. [39, 153].

Tap errors incurred during comb shaping are induced by the instability of the micro-combs and also the accuracy of the waveshapers. Instability of the micro-combs also results in power variations of the comb lines. A real-time feedback control path was used in Ref. [153] to reduce the related tap errors. The comb line powers were first detected by an optical spectrum analyser (OSA) and then compared with the ideal tap weights, generating an error signal that was fed back into the waveshaper to calibrate the system and achieve accurate comb shaping. In addition, mode-locking methods adopted in the generation of stable and coherent micro-combs can also be used here. Like the power dynamic range of optical amplifiers, a high shaping accuracy for the waveshapers is desired since it can also lead to decreased deviations and improved performances in practical processing.

## V. CONCLUSIONS

Micro-combs represent an exceptionally active field of research that has only developed in the last ten years. This young and fast-growing field has its roots in the pioneering development of micro-cavity technologies, starting in the late 1980s and experiencing significant advances in the last ten years, both for bulk and integrated resonators. We review the development of micro-comb based RF photonics and discuss the strong potential, as well as the challenges, that remain in this field. We introduce the physics underlying micro-comb generation and summarize a variety of photonic devices used for micro-comb generation. We review micro-comb based RF photonic devices, including RF photonic oscillators for microwave and MMW generation, RF photonic TTDLs for PAAs, RF photonic filters, and multi-functional RF photonic processors. Micro-combs bring a new generation of compact, low-cost, and highly efficient sources to the RF photonics community and offer enormous possibilities towards achieving high-performance RF signal generation, filtering, and processing.

**ACKNOWLEDGMENTS**

This work was supported by the Australian Research Council Discovery Projects Program (No. DP150104327). RM acknowledges support by Natural Sciences and Engineering Research Council of Canada (NSERC) through the Strategic, Discovery and Acceleration Grants Schemes, by the MESI PSR-SIIRI Initiative in Quebec, and by the Canada Research Chair Program. He also acknowledges additional support by the Government of the Russian Federation through the ITMO Fellowship and Professorship Program (grant 074-U 01) and by the 1000 Talents

Sichuan Program in China. Brent E. Little was supported by the Strategic Priority Research Program of the Chinese Academy of Sciences, Grant No. XDB24030000.

**REFERENCES**

[1] J. Capmany, and D. Novak, "Microwave photonics combines two worlds," *Nature Photonics,* vol. 1, no. 6, pp. 319-330, Jun, 2007.
[2] R. C. Williamson, and R. D. Esman, "RF photonics," *Journal of Lightwave Technology,* vol. 26, no. 9-12, pp. 1145-1153, May-Jun, 2008.
[3] J. P. Yao, "Microwave Photonics," *Journal of Lightwave Technology,* vol. 27, no. 1-4, pp. 314-335, Jan-Feb, 2009.
[4] A. J. Seeds, "Microwave photonics," *IEEE Transactions on Microwave Theory and Techniques,* vol. 50, no. 3, pp. 877-887, Mar, 2002.
[5] D. J. Jones, S. A. Diddams, J. K. Ranka, A. Stentz, R. S. Windeler, J. L. Hall, and S. T. Cundiff, "Carrier-envelope phase control of femtosecond mode-locked lasers and direct optical frequency synthesis," *Science,* vol. 288, no. 5466, pp. 635-639, Apr 28, 2000.
[6] S. A. Diddams, D. J. Jones, J. Ye, S. T. Cundiff, J. L. Hall, J. K. Ranka, R. S. Windeler, R. Holzwarth, T. Udem, and T. W. Hansch, "Direct link between microwave and optical frequencies with a 300 THz femtosecond laser comb," *Physical Review Letters,* vol. 84, no. 22, pp. 5102-5105, May 29, 2000.
[7] R. Holzwarth, T. Udem, T. W. Hansch, J. C. Knight, W. J. Wadsworth, and P. S. J. Russell, "Optical frequency synthesizer for precision spectroscopy," *Physical Review Letters,* vol. 85, no. 11, pp. 2264-2267, Sep 11, 2000.
[8] T. Udem, R. Holzwarth, and T. W. Hansch, "Optical frequency metrology," *Nature,* vol. 416, no. 6877, pp. 233-237, Mar 14, 2002.
[9] T. W. Hansch, "Passion for precision," *Chemphyschem,* vol. 7, no. 6, pp. 1170-+, Jun 12, 2006.
[10] I. Coddington, W. C. Swann, and N. R. Newbury, "Coherent multiheterodyne spectroscopy using stabilized optical frequency combs (vol 100, art no 013902, 2008)," *Physical Review Letters,* vol. 101, no. 4, Jul 25, 2008.
[11] B. Bernhardt, A. Ozawa, P. Jacquet, M. Jacquey, Y. Kobayashi, T. Udem, R. Holzwarth, G. Guelachvili, T. W. Hansch, and N. Picque, "Cavity-enhanced dual-comb spectroscopy," *Nature Photonics,* vol. 4, no. 1, pp. 55-57, Jan 10, 2010.
[12] F. Adler, M. J. Thorpe, K. C. Cossel, and J. Ye, "Cavity-Enhanced Direct Frequency Comb Spectroscopy: Technology and Applications," *Annual Review of Analytical Chemistry, Vol 3,* vol. 3, pp. 175-205, 2010.
[13] T. Ideguchi, A. Poisson, G. Guelachvili, N. Picque, and T. W. Hansch, "Adaptive real-time dual-comb spectroscopy," *Nature Communications,* vol. 5, Feb, 2014.
[14] J. M. Lukens, D. E. Leaird, and A. M. Weiner, "A temporal cloak at telecommunication data rate," *Nature,* vol. 498, no. 7453, pp. 205-208, Jun 13, 2013.
[15] F. Monticone, C. Argyropoulos, and A. Alu, "Multilayered Plasmonic Covers for Comblike Scattering Response and Optical Tagging," *Physical Review Letters,* vol. 110, no. 11, Mar 12, 2013.
[16] N. C. Menicucci, S. T. Flammia, and O. Pfister, "One-way quantum computing in the optical frequency comb," *Physical Review Letters,* vol. 101, no. 13, Sep 26, 2008.
[17] O. Pinel, P. Jian, R. M. de Araujo, J. X. Feng, B. Chalopin, C. Fabre, and N. Treps, "Generation and Characterization of Multimode Quantum Frequency Combs," *Physical Review Letters,* vol. 108, no. 8, Feb 23, 2012.
[18] M. Chen, N. C. Menicucci, and O. Pfister, "Experimental Realization of Multipartite Entanglement of 60 Modes of a Quantum Optical Frequency Comb," *Physical Review Letters,* vol. 112, no. 12, Mar 26, 2014.
[19] D. Hillerkuss, R. Schmogrow, T. Schellinger, M. Jordan, M. Winter, G. Huber, T. Vallaitis, R. Bonk, P. Kleinow, F. Frey, M. Roeger, S. Koenig, A. Ludwig, A. Marculescu, J. Li, M. Hoh, M. Dreschmann, J. Meyer, S. Ben Ezra, N. Narkiss, B. Nebendahl, F. Parmigiani, P. Petropoulos, B. Resan, A. Oehler, K. Weingarten, T. Ellermeyer, J. Lutz, M. Moeller, M. Huebner, J. Becker, C. Koos, W. Freude, and J. Leuthold, "26 Tbit s(-1) line-rate super-channel transmission utilizing all-optical fast Fourier transform processing," *Nature Photonics,* vol. 5, no. 6, pp. 364-371, Jun, 2011.
[20] D. Hillerkuss, R. Schmogrow, M. Meyer, S. Wolf, M. Jordan, P. Kleinow, N. Lindenmann, P. C. Schindler, A. Melikyan, X. Yang, S. Ben-Ezra, B. Nebendahl, M. Dreschmann, J. Meyer, F. Parmigiani, P. Petropoulos, B. Resan, A. Oehler, K. Weingarten, L. Altenhain, T. Ellermeyer, M. Moeller, M. Huebner, J. Becker, C. Koos, W. Freude, and J. Leuthold, "Single-Laser 32.5 Tbit/s Nyquist WDM Transmission," *Journal of Optical Communications and Networking,* vol. 4, no. 10, pp. 715-723, Oct, 2012.

[21] S. T. Cundiff, and A. M. Weiner, "Optical arbitrary waveform generation," *Nature Photonics,* vol. 4, no. 11, pp. 760-766, Nov, 2010.
[22] T. M. Fortier, M. S. Kirchner, F. Quinlan, J. Taylor, J. C. Bergquist, T. Rosenband, N. Lemke, A. Ludlow, Y. Jiang, C. W. Oates, and S. A. Diddams, "Generation of ultrastable microwaves via optical frequency division," *Nature Photonics,* vol. 5, no. 7, pp. 425-429, Jul, 2011.
[23] V. R. Supradeepa, C. M. Long, R. Wu, F. Ferdous, E. Hamidi, D. E. Leaird, and A. M. Weiner, "Comb-based radiofrequency photonic filters with rapid tunability and high selectivity," *Nature Photonics,* vol. 6, no. 3, pp. 186-194, Mar, 2012.
[24] V. Torres-Company, and A. M. Weiner, "Optical frequency comb technology for ultra-broadband radio-frequency photonics," *Laser & Photonics Reviews,* vol. 8, no. 3, pp. 368-393, May, 2014.
[25] S. T. Cundiff, "Metrology - New generation of combs," *Nature,* vol. 450, no. 7173, pp. 1175-1176, Dec 20, 2007.
[26] V. Brasch, M. Geiselmann, T. Herr, G. Lihachev, M. H. P. Pfeiffer, M. L. Gorodetsky, and T. J. Kippenberg, "Photonic chip-based optical frequency comb using soliton Cherenkov radiation," *Science,* vol. 351, no. 6271, pp. 357-360, Jan 22, 2016.
[27] S. B. Papp, K. Beha, P. Del'Haye, F. Quinlan, H. Lee, K. J. Vahala, and S. A. Diddams, "Microresonator frequency comb optical clock," *Optica,* vol. 1, no. 1, pp. 10-14, Jul 22, 2014.
[28] M.Peccianti, A.Pasquazi, Y.Park, B.E Little, S.Chu, D.J Moss, and R.Morandotti, "Demonstration of an ultrafast nonlinear microcavity modelocked laser", Nature Communications **3** 765 (2012). DOI:10.1038/ncomms1762
[29] K. Saha, Y. Okawachi, B. Shim, J. S. Levy, R. Salem, A. R. Johnson, M. A. Foster, M. R. E. Lamont, M. Lipson, and A. L. Gaeta, "Modelocking and femtosecond pulse generation in chip-based frequency combs," *Optics Express,* vol. 21, no. 1, pp. 1335-1343, Jan 14, 2013.
[30] S. W. Huang, H. Zhou, J. Yang, J. F. McMillan, A. Matsko, M. Yu, D. L. Kwong, L. Maleki, and C. W. Wong, "Mode-Locked Ultrashort Pulse Generation from On-Chip Normal Dispersion Microresonators," *Physical Review Letters,* vol. 114, no. 5, Feb 4, 2015.
[31] M. Kues, C. Reimer, B. Wetzel, P. Roztocki, B. E. Little, S. T. Chu, T. Hansson, E. A. Viktorov, D. J. Moss, and R. Morandotti, "Passively mode-locked laser with an ultra-narrow spectral width," *Nature Photonics,* vol. **11**, no. 3, pp. 159-162, (2017).
[32] J. Pfeifle, V. Brasch, M. Lauermann, Y. M. Yu, D. Wegner, T. Herr, K. Hartinger, P. Schindler, J. S. Li, D. Hillerkuss, R. Schmogrow, C. Weimann, R. Holzwarth, W. Freude, J. Leuthold, T. J. Kippenberg, and C. Koos, "Coherent terabit communications with microresonator Kerr frequency combs," *Nature Photonics,* vol. 8, no. 5, pp. 375-380, May, 2014.
[33] C. J. Bao, P. C. Liao, A. Kordts, M. Karpov, M. H. P. Pfeiffer, L. Zhang, Y. Yan, G. D. Xie, Y. W. Cao, A. Almaiman, M. Ziyadi, L. Li, Z. Zhao, A. Mohajerin-Ariaei, S. R. Wilkinson, M. Tur, M. M. Fejer, T. J. Kippenberg, and A. E. Willner, "Demonstration of optical multicasting using Kerr frequency comb lines," *Optics Letters,* vol. 41, no. 16, pp. 3876-3879, Aug 15, 2016.
[34] P. C. Liao, C. J. Bao, A. Kordts, M. Karpov, M. H. P. Pfeiffer, L. Zhang, Y. W. Cao, A. Almaiman, A. Mohajerin-Ariaei, M. Tur, M. M. Fejer, T. J. Kippenberg, and A. E. Willner, "Pump-linewidth-tolerant wavelength multicasting using soliton Kerr frequency combs," *Optics Letters,* vol. 42, no. 16, pp. 3177-3180, Aug 15, 2017.
[35] P. Marin-Palomo, J. N. Kemal, M. Karpov, A. Kordts, J. Pfeifle, M. H. P. Pfeiffer, P. Trocha, S. Wolf, V. Brasch, M. H. Anderson, R. Rosenberger, K. Vijayan, W. Freude, T. J. Kippenberg, and C. Koos, "Microresonator-based solitons for massively parallel coherent optical communications," *Nature,* vol. 546, no. 7657, pp. 274-+, Jun 8, 2017.
[36] P. Del'Haye, T. Herr, E. Gavartin, M. L. Gorodetsky, R. Holzwarth, and T. J. Kippenberg, "Octave Spanning Tunable Frequency Comb from a Microresonator," *Physical Review Letters,* vol. 107, no. 6, Aug 1, 2011.
[37] J. Li, H. Lee, T. Chen, and K. J. Vahala, "Low-Pump-Power, Low-Phase-Noise, and Microwave to Millimeter-Wave Repetition Rate Operation in Microcombs," *Physical Review Letters,* vol. 109, no. 23, Dec 4, 2012.
[38] W. Liang, D. Eliyahu, V. S. Ilchenko, A. A. Savchenkov, A. B. Matsko, D. Seidel, and L. Maleki, "High spectral purity Kerr frequency comb radio frequency photonic oscillator," *Nature Communications,* vol. 6, Aug, 2015.
[39] T. G. Nguyen, M. Shoeiby, S. T. Chu, B. E. Little, R. Morandotti, A. Mitchell, and D. J. Moss, "Integrated frequency comb source based Hilbert transformer for wideband microwave photonic phase analysis," *Optics Express,* vol. 23, no. 17, pp. 22087-22097, (2015).
[40] C. Reimer, M. Kues, P. Roztocki, B. Wetzel, F. Grazioso, B. E. Little, S. T. Chu, T. Johnston, Y. Bromberg, L. Caspani, D. J. Moss, and R. Morandotti, "Generation of multiphoton entangled quantum states by means of integrated frequency combs," *Science,* vol. **351**, no. 6278, pp. 1176-1180, (2016).

[41] M. Kues, et al., "On-chip generation of high-dimensional entangled quantum states and their coherent control", Nature **546** (7660) 622-626 (2017).
[42] T. J. Kippenberg, R. Holzwarth, and S. A. Diddams, "Microresonator-Based Optical Frequency Combs," *Science,* vol. **332**, no. 6029, pp. 555-559, Apr 29, 2011.
[43] Y. K. Chembo, "Kerr optical frequency combs: theory, applications and perspectives," *Nanophotonics,* vol. 5, no. 2, pp. 214-230, Jun, 2016.
[44] X. X. Xue, M. H. Qi, and A. M. Weiner, "Normal-dispersion microresonator Kerr frequency combs," *Nanophotonics,* vol. 5, no. 2, pp. 244-262, Jun, 2016.
[45] A. A. Savchenkov, A. B. Matsko, and L. Maleki, "On Frequency Combs in Monolithic Resonators," *Nanophotonics,* vol. 5, no. 2, pp. 363-391, Jun, 2016.
[46] T. Hansson, and S. Wabnitz, "Dynamics of microresonator frequency comb generation: models and stability," *Nanophotonics,* vol. 5, no. 2, pp. 231-243, Jun, 2016.
[47] X. X. Xue, and A. M. Weiner, "Microwave photonics connected with microresonator frequency combs," *Frontiers of Optoelectronics,* vol. 9, no. 2, pp. 238-248, Jun, 2016.
[48] P. Del'Haye, A. Schliesser, O. Arcizet, T. Wilken, R. Holzwarth, and T. J. Kippenberg, "Optical frequency comb generation from a monolithic microresonator," *Nature,* vol. 450, no. 7173, pp. 1214-1217, Dec 20, 2007.
[49] L. Razzari, D. Duchesne, M. Ferrera, R. Morandotti, S. Chu, B. E. Little, and D. J. Moss, "CMOS-compatible integrated optical hyper-parametric oscillator," *Nature Photonics,* vol. **4**, no. 1, pp. 41-45, (2010).
[50] J. S. Levy, A. Gondarenko, M. A. Foster, A. C. Turner-Foster, A. L. Gaeta, and M. Lipson, "CMOS-compatible multiple-wavelength oscillator for on-chip optical interconnects," *Nature Photonics,* vol. **4**, no. 1, pp. 37-40, Jan 10, 2010.
[51] A. Pasquazi, M. Pecciantia, L. Razzari, David J. Moss, S. Coen, M. Erkintalo, Y.K. Chembo, T. Hansson, S. Wabnitz, P. del'Haye, X. Xue, A.M. Weiner, R. Morandotti, "Micro-Combs: A Novel Generation of Optical Sources", **Physics Reports 729** 1–81 (2018). DOI:10.1016/j.physrep.2017.08.004.
[52] D. J. Moss, R. Morandotti, A. L. Gaeta, and M. Lipson, "New CMOS-compatible platforms based on silicon nitride and Hydex for nonlinear optics," *Nature Photonics,* vol. **7**, no. 8, pp. 597-607, (2013).
[53] A. B. Matsko, and V. S. Ilchenko, "Optical resonators with whispering-gallery modes - Part I: Basics," *IEEE Journal of Selected Topics in Quantum Electronics,* vol. 12, no. 1, pp. 3-14, Jan-Feb, 2006.
[54] V. S. Ilchenko, and A. B. Matsko, "Optical resonators with whispering-gallery modes - Part II: Applications," *IEEE Journal of Selected Topics in Quantum Electronics,* vol. 12, no. 1, pp. 15-32, Jan-Feb, 2006.
[55] I. H. Agha, Y. Okawachi, and A. L. Gaeta, "Theoretical and experimental investigation of broadband cascaded four-wave mixing in high-Q microspheres," *Optics Express,* vol. 17, no. 18, pp. 16209-16215, Aug 31, 2009.
[56] S. B. Papp, P. Del'Haye, and S. A. Diddams, "Mechanical Control of a Microrod-Resonator Optical Frequency Comb," *Physical Review X,* vol. 3, no. 3, Jul 8, 2013.
[57] T. Herr, V. Brasch, J. D. Jost, C. Y. Wang, N. M. Kondratiev, M. L. Gorodetsky, and T. J. Kippenberg, "Temporal solitons in optical microresonators," *Nature Photonics,* vol. 8, no. 2, pp. 145-152, Feb, 2014.
[58] "Extending opportunities," *Nature Photonics,* vol. 6, no. 7, pp. 407-407, Jul, 2012.
[59] A. G. Griffith, R. K. W. Lau, J. Cardenas, Y. Okawachi, A. Mohanty, R. Fain, Y. H. D. Lee, M. J. Yu, C. T. Phare, C. B. Poitras, A. L. Gaeta, and M. Lipson, "Silicon-chip mid-infrared frequency comb generation," *Nature Communications,* vol. 6, Feb, 2015.
[60] H. Jung, R. Stoll, X. Guo, D. Fischer, and H. X. Tang, "Green, red, and IR frequency comb line generation from single IR pump in AlN microring resonator," *Optica,* vol. 1, no. 6, pp. 396-399, Dec 20, 2014.
[61] B. J. M. Hausmann, I. Bulu, V. Venkataraman, P. Deotare, and M. Loncar, "Diamond nonlinear photonics," *Nature Photonics,* vol. 8, no. 5, pp. 369-374, May, 2014.
[62] W. Liang, A. A. Savchenkov, A. B. Matsko, V. S. Ilchenko, D. Seidel, and L. Maleki, "Generation of near-infrared frequency combs from a MgF2 whispering gallery mode resonator," *Optics Letters,* vol. 36, no. 12, pp. 2290-2292, Jun 15, 2011.
[63] C. Y. Wang, T. Herr, P. Del'Haye, A. Schliesser, J. Hofer, R. Holzwarth, T. W. Hansch, N. Picque, and T. J. Kippenberg, "Mid-infrared optical frequency combs at 2.5 mu m based on crystalline microresonators," *Nature Communications,* vol. 4, Jan, 2013.
[64] W. Liang, A. A. Savchenkov, Z. D. Xie, J. F. McMillan, J. Burkhart, V. S. Ilchenko, C. W. Wong, A. B. Matsko, and L. Maleki, "Miniature multioctave light source based on a monolithic microcavity," *Optica,* vol. 2, no. 1, pp. 40-47, Jan 20, 2015.
[65] I. S. Grudinin, N. Yu, and L. Maleki, "Generation of optical frequency combs with a CaF2 resonator," *Optics Letters,* vol. 34, no. 7, pp. 878-880, Apr 1, 2009.
[66] A. A. Savchenkov, A. B. Matsko, W. Liang, V. S. Ilchenko, D. Seidel, and L. Maleki, "Kerr combs with

selectable central frequency," *Nature Photonics,* vol. 5, no. 5, pp. 293-296, May, 2011.
[67] W. Q. Wang, S. T. Chu, B. E. Little, A. Pasquazi, Y. S. Wang, L. R. Wang, W. F. Zhang, L. Wang, X. H. Hu, G. X. Wang, H. Hu, Y. L. Su, F. T. Li, Y. S. Liu, and W. Zhao, "Dual-pump Kerr Micro-cavity Optical Frequency Comb with varying FSR spacing," *Scientific Reports,* vol. 6, Jun 24, 2016.
[68] Y. Okawachi, K. Saha, J. S. Levy, Y. H. Wen, M. Lipson, and A. L. Gaeta, "Octave-spanning frequency comb generation in a silicon nitride chip," *Optics Letters,* vol. 36, no. 17, pp. 3398-3400, Sep 1, 2011.
[69] L. R. Wang, L. Chang, N. Volet, M. H. P. Pfeiffer, M. Zervas, H. R. Guo, T. J. Kippenberg, and J. E. Bowers, "Frequency comb generation in the green using silicon nitride microresonators," *Laser & Photonics Reviews,* vol. 10, no. 4, pp. 631-638, Jul, 2016.
[70] Y. Xuan, Y. Liu, L. T. Varghese, A. J. Metcalf, X. X. Xue, P. H. Wang, K. Han, J. A. Jaramillo-Villegas, A. Al Noman, C. Wang, S. Kim, M. Teng, Y. J. Lee, B. Niu, L. Fan, J. Wang, D. E. Leaird, A. M. Weiner, and M. H. Qi, "High-Q silicon nitride microresonators exhibiting low-power frequency comb initiation," *Optica,* vol. 3, no. 11, pp. 1171-1180, Nov 20, 2016.
[71] H. Jung, C. Xiong, K. Y. Fong, X. F. Zhang, and H. X. Tang, "Optical frequency comb generation from aluminum nitride microring resonator," *Optics Letters,* vol. 38, no. 15, pp. 2810-2813, Aug 1, 2013.
[72] Y. K. Chembo, D. V. Strekalov, and N. Yu, "Spectrum and Dynamics of Optical Frequency Combs Generated with Monolithic Whispering Gallery Mode Resonators," *Physical Review Letters,* vol. 104, no. 10, Mar 12, 2010.
[73] D. Duchesne, M. Ferrera, L. Razzari, R. Morandotti, B. E. Little, S. T. Chu, and D. J. Moss, "Efficient self-phase modulation in low loss, high index doped silica glass integrated waveguides," *Optics Express,* vol. **17**, no. 3, pp. 1865-1870, (2009).
[74] J. Leuthold, C. Koos, and W. Freude, "Nonlinear silicon photonics," *Nature Photonics,* vol. 4, no. 8, pp. 535-544, Aug, 2010.
[75] D. K. Armani, T. J. Kippenberg, S. M. Spillane, and K. J. Vahala, "Ultra-high-Q toroid microcavity on a chip," *Nature,* vol. 421, no. 6926, pp. 925-928, Feb 27, 2003.
[76] X. M. Zhang, H. S. Choi, and A. M. Armani, "Ultimate quality factor of silica microtoroid resonant cavities," *Applied Physics Letters,* vol. 96, no. 15, Apr 12, 2010.
[77] M. L. Gorodetsky, A. A. Savchenkov, and V. S. Ilchenko, "Ultimate Q of optical microsphere resonators," *Optics Letters,* vol. 21, no. 7, pp. 453-455, Apr 1, 1996.
[78] A. Chiasera, Y. Dumeige, P. Feron, M. Ferrari, Y. Jestin, G. N. Conti, S. Pelli, S. Soria, and G. C. Righini, "Spherical whispering-gallery-mode microresonators," *Laser & Photonics Reviews,* vol. 4, no. 3, pp. 457-482, May, 2010.
[79] P. Del'Haye, O. Arcizet, A. Schliesser, R. Holzwarth, and T. J. Kippenberg, "Full stabilization of a microresonator-based optical frequency comb," *Physical Review Letters,* vol. 101, no. 5, Aug 1, 2008.
[80] I. S. Grudinin, A. B. Matsko, A. A. Savchenkov, D. Strekalov, V. S. Ilchenko, and L. Maleki, "Ultra high Q crystalline microcavities," *Optics Communications,* vol. 265, no. 1, pp. 33-38, Sep 1, 2006.
[81] C. Lecaplain, C. Javerzac-Galy, M. L. Gorodetsky, and T. J. Kippenberg, "Mid-infrared ultra-high-Q resonators based on fluoride crystalline materials," *Nature Communications,* vol. 7, Nov 21, 2016.
[82] A. A. Savchenkov, A. B. Matsko, V. S. Ilchenko, and L. Maleki, "Optical resonators with ten million finesse," *Optics Express,* vol. 15, no. 11, pp. 6768-6773, May 28, 2007.
[83] G. P. Lin, S. Diallo, R. Henriet, M. Jacquot, and Y. K. Chembo, "Barium fluoride whispering-gallery-mode disk-resonator with one billion quality-factor," *Optics Letters,* vol. 39, no. 20, pp. 6009-6012, Oct 15, 2014.
[84] R. Henriet, G. P. Lin, A. Coillet, M. Jacquot, L. Furfaro, L. Larger, and Y. K. Chembo, "Kerr optical frequency comb generation in strontium fluoride whispering-gallery mode resonators with billion quality factor," *Optics Letters,* vol. 40, no. 7, pp. 1567-1570, Apr 1, 2015.
[85] R. Soref, "The past, present, and future of silicon photonics," *Ieee Journal of Selected Topics in Quantum Electronics,* vol. 12, no. 6, pp. 1678-1687, Nov-Dec, 2006.
[86] Q. F. Xu, D. Fattal, and R. G. Beausoleil, "Silicon microring resonators with 1.5-mu m radius," *Optics Express,* vol. 16, no. 6, pp. 4309-4315, Mar 17, 2008.
[87] J. Y. Wu, B. Y. Liu, J. Z. Peng, J. M. Mao, X. H. Jiang, C. Y. Qiu, C. Tremblay, and Y. K. Su, "On-Chip Tunable Second-Order Differential-Equation Solver Based on a Silicon Photonic Mode-Split Microresonator," *Journal of Lightwave Technology,* vol. 33, no. 17, pp. 3542-3549, Sep 1, 2015.
[88] J.Wu, T.Moein, X.Xu, G.Ren, A. Mitchell, and D.J. Moss, "Micro-ring resonator quality factor enhancement via an integrated Fabry-Perot cavity", Applied Physics Letters (APL) Photonics **2** 056103 (2017).
[89] K. Ikeda, R. E. Saperstein, N. Alic, and Y. Fainman, "Thermal and Kerr nonlinear properties of plasma-deposited silicon nitride/silicon dioxide waveguides," *Optics Express,* vol. 16, no. 17, pp. 12987-12994, Aug 18, 2008.
[90] K. Luke, A. Dutt, C. B. Poitras, and M. Lipson, "Overcoming Si3N4 film stress limitations for high

quality factor ring resonators," *Optics Express,* vol. 21, no. 19, pp. 22829-22833, Sep 23, 2013.
[91] M. H. P. Pfeiffer, A. Kordts, V. Brasch, M. Zervas, M. Geiselmann, J. D. Jost, and T. J. Kippenberg, "Photonic Damascene process for integrated high-Q microresonator based nonlinear photonics," *Optica,* vol. 3, no. 1, pp. 20-25, Jan 20, 2016.
[92] Y. Liu, Y. Xuan, X. X. Xue, P. H. Wang, S. Chen, A. J. Metcalf, J. Wang, D. E. Leaird, M. H. Qi, and A. M. Weiner, "Investigation of mode coupling in normal-dispersion silicon nitride microresonators for Kerr frequency comb generation," *Optica,* vol. 1, no. 3, pp. 137-144, Sep 20, 2014.
[93] X. X. Xue, Y. Xuan, Y. Liu, P. H. Wang, S. Chen, J. Wang, D. E. Leaird, M. H. Qi, and A. M. Weiner, "Mode-locked dark pulse Kerr combs in normal-dispersion microresonators," *Nature Photonics,* vol. 9, no. 9, pp. 594-+, Sep, 2015.
[94] X. X. Xue, Y. Xuan, P. H. Wang, Y. Liu, D. E. Leaird, M. H. Qi, and A. M. Weiner, "Normal-dispersion microcombs enabled by controllable mode interactions," *Laser & Photonics Reviews,* vol. 9, no. 4, pp. L23-L28, Jul, 2015.
[95] S. Kim, K. Han, C. Wang, J. A. Jaramillo-Villegas, X. X. Xue, C. Y. Bao, Y. Xuan, D. E. Leaird, A. M. Weiner, and M. H. Qi, "Dispersion engineering and frequency comb generation in thin silicon nitride concentric microresonators," *Nature Communications,* vol. 8, Aug 29, 2017.
[96] X. X. Xue, F. Leo, Y. Xuan, J. A. Jaramillo-Villegas, P. H. Wang, D. E. Leaird, M. Erkintalo, M. H. Qi, and A. M. Weiner, "Second-harmonic-assisted four-wave mixing in chip-based microresonator frequency comb generation," *Light-Science & Applications,* vol. 6, Apr, 2017.
[97] S. Miller, K. Luke, Y. Okawachi, J. Cardenas, A. L. Gaeta, and M. Lipson, "On-chip frequency comb generation at visible wavelengths via simultaneous second- and third-order optical nonlinearities," *Optics Express,* vol. 22, no. 22, pp. 26517-26525, Nov 3, 2014.
[98] B. E. Little, S. T. Chu, P. P. Absil, J. V. Hryniewicz, F. G. Johnson, E. Seiferth, D. Gill, V. Van, O. King, and M. Trakalo, "Very high-order microring resonator filters for WDM applications," *Ieee Photonics Technology Letters,* vol. 16, no. 10, pp. 2263-2265, Oct, 2004.
[99] M. Ferrera, L. Razzari, D. Duchesne, R. Morandotti, Z. Yang, M. Liscidini, J. E. Sipe, S. Chu, B. E. Little, and D. J. Moss, "Low-power continuous-wave nonlinear optics in doped silica glass integrated waveguide structures," *Nature Photonics,* vol. **2**, no. 12, pp. 737-740, (2008).
[100] M. Ferrera, et al., "Ultra-Fast Integrated All-Optical Integrator", Nature Communications **1** Article 29 (2010). DOI:10.1038/ncomms1028
[101] T. J. Kippenberg, S. M. Spillane, and K. J. Vahala, "Kerr-nonlinearity optical parametric oscillation in an ultrahigh-Q toroid microcavity," *Physical Review Letters,* vol. 93, no. 8, Aug 20, 2004.
[102] A. A. Savchenkov, A. B. Matsko, D. Strekalov, M. Mohageg, V. S. Ilchenko, and L. Maleki, "Low threshold optical oscillations in a whispering gallery mode CaF2 resonator," *Physical Review Letters,* vol. 93, no. 24, Dec 10, 2004.
[103] L. Maleki, "SOURCES The optoelectronic oscillator," *Nature Photonics,* vol. 5, no. 12, pp. 728-730, Dec, 2011.
[104] K. Tai, A. Hasegawa, and A. Tomita, "Observation of Modulational Instability in Optical Fibers," *Physical Review Letters,* vol. 56, no. 2, pp. 135-138, Jan 13, 1986.
[105] A. B. Matsko, W. Liang, A. A. Savchenkov, and L. Maleki, "Chaotic dynamics of frequency combs generated with continuously pumped nonlinear microresonators," *Optics Letters,* vol. 38, no. 4, pp. 525-527, Feb 15, 2013.
[106] M. J. Yu, J. K. Jang, Y. Okawachi, A. G. Griffith, K. Luke, S. A. Miller, X. C. Ji, M. Lipson, and A. L. Gaeta, "Breather soliton dynamics in microresonators," *Nature Communications,* vol. 8, Feb 24, 2017.
[107] H. Guo, M. Karpov, E. Lucas, A. Kordts, M. H. P. Pfeiffer, V. Brasch, G. Lihachev, V. E. Lobanov, M. L. Gorodetsky, and T. J. Kippenberg, "Universal dynamics and deterministic switching of dissipative Kerr solitons in optical microresonators," *Nature Physics,* vol. 13, no. 1, pp. 94-102, Jan, 2017.
[108] Y. K. Chembo, and N. Yu, "Modal expansion approach to optical-frequency-comb generation with monolithic whispering-gallery-mode resonators," *Physical Review A,* vol. 82, no. 3, Sep 7, 2010.
[109] A. B. Matsko, A. A. Savchenkov, W. Liang, V. S. Ilchenko, D. Seidel, and L. Maleki, "Mode-locked Kerr frequency combs," *Optics Letters,* vol. 36, no. 15, pp. 2845-2847, Aug 1, 2011.
[110] S. Coen, H. G. Randle, T. Sylvestre, and M. Erkintalo, "Modeling of octave-spanning Kerr frequency combs using a generalized mean-field Lugiato-Lefever model," *Optics Letters,* vol. 38, no. 1, pp. 37-39, Jan 1, 2013.
[111] Y. K. Chembo, and C. R. Menyuk, "Spatiotemporal Lugiato-Lefever formalism for Kerr-comb generation in whispering-gallery-mode resonators," *Physical Review A,* vol. 87, no. 5, May 31, 2013.
[112] C. Godey, I. V. Balakireva, A. Coillet, and Y. K. Chembo, "Stability analysis of the spatiotemporal Lugiato-Lefever model for Kerr optical frequency combs in the anomalous and normal dispersion regimes," *Physical Review A,* vol. 89, no. 6, Jun 16, 2014.
[113] K. Ikeda, "Multiple-Valued Stationary State and Its Instability of the Transmitted Light by a Ring Cavity

System," *Optics Communications,* vol. 30, no. 2, pp. 257-261, 1979.

[114] M. Erkintalo, and S. Coen, "Coherence properties of Kerr frequency combs," *Optics Letters,* vol. 39, no. 2, pp. 283-286, Jan 15, 2014.

[115] S. Coen, and M. Erkintalo, "Universal scaling laws of Kerr frequency combs," *Optics Letters,* vol. 38, no. 11, pp. 1790-1792, Jun 1, 2013.

[116] H. M. Gibbs, S. L. Mccall, and T. N. C. Venkatesan, "Differential Gain and Bistability Using a Sodium-Filled Fabry-Perot-Interferometer," *Physical Review Letters,* vol. 36, no. 19, pp. 1135-1138, 1976.

[117] T. Hansson, D. Modotto, and S. Wabnitz, "Dynamics of the modulational instability in microresonator frequency combs," *Physical Review A,* vol. 88, no. 2, Aug 12, 2013.

[118] A. Coillet, J. Dudley, G. Genty, L. Larger, and Y. K. Chembo, "Optical rogue waves in whispering-gallery-mode resonators," *Physical Review A,* vol. 89, no. 1, Jan 24, 2014.

[119] D. R. Solli, C. Ropers, P. Koonath, and B. Jalali, "Optical rogue waves," *Nature,* vol. 450, no. 7172, pp. 1054-U7, Dec 13, 2007.

[120] V. E. Lobanov, G. Lihachev, T. J. Kippenberg, and M. L. Gorodetsky, "Frequency combs and platicons in optical microresonators with normal GVD," *Optics Express,* vol. 23, no. 6, pp. 7713-7721, Mar 23, 2015.

[121] A. Coillet, I. Balakireva, R. Henriet, K. Saleh, L. Larger, J. M. Dudley, C. R. Menyuk, and Y. K. Chembo, "Azimuthal Turing Patterns, Bright and Dark Cavity Solitons in Kerr Combs Generated With Whispering-Gallery-Mode Resonators," *Ieee Photonics Journal,* vol. 5, no. 4, Aug, 2013.

[122] C. J. Bao, P. C. Liao, L. Zhang, Y. Yan, Y. W. Cao, G. D. Xie, A. Mohajerin-Ariaei, L. Li, M. Ziyadi, A. Almaiman, L. C. Kimerling, J. Michel, and A. E. Willner, "Effect of a breather soliton in Kerr frequency combs on optical communication systems," *Optics Letters,* vol. 41, no. 8, pp. 1764-1767, Apr 15, 2016.

[123] M. Ferrera, C. Reimer, A. Pasquazi, M. Peccianti, M. Clerici, L. Caspani, S. T. Chu, B. E. Little, R. Morandotti, and D. J. Moss, "CMOS compatible integrated all-optical radio frequency spectrum analyzer," *Optics Express,* vol. **22**, no. 18, pp. 21488-21498, (2014).

[124] B. Corcoran, T. D. Vo, M. D. Pelusi, C. Monat, D. X. Xu, A. Densmore, R. B. Ma, S. Janz, D. J. Moss, and B. J. Eggleton, "Silicon nanowire based radio-frequency spectrum analyzer," *Optics Express,* vol. **18**, no. 19, pp. 20190-20200, (2010).

[125] M. Pelusi, F. Luan, T. D. Vo, M. R. E. Lamont, S. J. Madden, D. A. Bulla, D. Y. Choi, B. Luther-Davies, and B. J. Eggleton, "Photonic-chip-based radio-frequency spectrum analyser with terahertz bandwidth," *Nature Photonics,* vol. 3, no. 3, pp. 139-143, Mar, 2009.

[126] J. Y. Wu, J. Z. Peng, B. Y. Liu, T. Pan, H. Y. Zhou, J. M. Mao, Y. X. Yang, C. Y. Qiu, and Y. K. Su, "Passive silicon photonic devices for microwave photonic signal processing," *Optics Communications,* vol. 373, pp. 44-52, Aug 15, 2016.

[127] Z. Jia, J. Yu, G. Ellinas, and G. K. Chang, "Key enabling technologies for optical-wireless networks: Optical millimeter-wave generation, wavelength reuse, and architecture," *Journal of Lightwave Technology,* vol. 25, no. 11, pp. 3452-3471, Nov, 2007.

[128] J. Li, X. Yi, H. Lee, S. A. Diddams, and K. J. Vahala, "Electro-optical frequency division and stable microwave synthesis," *Science,* vol. 345, no. 6194, pp. 309-313, Jul 18, 2014.

[129] A. Pasquazi, M. Peccianti, B. E. Little, S. T. Chu, D. J. Moss, and R. Morandotti, "Stable, dual mode, high repetition rate mode-locked laser based on a microring resonator," *Optics Express,* vol. 20, no. 24, pp. 27355-27362, (2012).

[130] S. B. Papp, and S. A. Diddams, "Spectral and temporal characterization of a fused-quartz-microresonator optical frequency comb," *Physical Review A,* vol. 84, no. 5, Nov 17, 2011.

[131] A. A. Savchenkov, E. Rubiola, A. B. Matsko, V. S. Ilchenko, and L. Maleki, "Phase noise of whispering gallery photonic hyper-parametric microwave oscillators," *Optics Express,* vol. 16, no. 6, pp. 4130-4144, Mar 17, 2008.

[132] A. B. Matsko, and L. Maleki, "On timing jitter of mode locked Kerr frequency combs," *Optics Express,* vol. 21, no. 23, pp. 28862-28876, Nov 18, 2013.

[133] W. Liang, V. S. Ilchenko, D. Eliyahu, A. A. Savchenkov, A. B. Matsko, D. Seidel, and L. Maleki, "Ultralow noise miniature external cavity semiconductor laser," *Nature Communications,* vol. 6, Jun, 2015.

[134] A. B. Matsko, and L. Maleki, "Noise conversion in Kerr comb RF photonic oscillators," *Journal of the Optical Society of America B-Optical Physics,* vol. 32, no. 2, pp. 232-240, Feb, 2015.

[135] K. Saleh, and Y. K. Chembo, "On the phase noise performance of microwave and millimeter-wave signals generated with versatile Kerr optical frequency combs," *Optics Express,* vol. 24, no. 22, pp. 25043-25056, Oct 31, 2016.

[136] D. H. Yang, and W. P. Lin, "Phased-array beam steering using optical true time delay technique," *Optics Communications,* vol. 350, pp. 90-96, Sep 1, 2015.

[137] X. W. Ye, F. Z. Zhang, and S. L. Pan, "Optical true time delay unit for multi-beamforming," *Optics*

*Express,* vol. 23, no. 8, pp. 10002-10008, Apr 20, 2015.
[138] Y. Q. Liu, J. P. Yao, and J. L. Yang, "Wideband true-time-delay unit for phased array beamforming using discrete-chirped fiber grating prism," *Optics Communications,* vol. 207, no. 1-6, pp. 177-187, Jun 15, 2002.
[139] S. Chin, L. Thevenaz, J. Sancho, S. Sales, J. Capmany, P. Berger, J. Bourderionnet, and D. Dolfi, "Broadband true time delay for microwave signal processing, using slow light based on stimulated Brillouin scattering in optical fibers," *Optics Express,* vol. 18, no. 21, pp. 22599-22613, Oct 11, 2010.
[140] J. J. Zhang, and J. P. Yao, "Photonic True-Time Delay Beamforming Using a Switch-Controlled Wavelength-Dependent Recirculating Loop," *Journal of Lightwave Technology,* vol. 34, no. 16, pp. 3923-3929, Aug 15, 2016.
[141] M. Longbrake, "True Time-Delay Beamsteering for Radar," *Proceedings of the 2012 Ieee National Aerospace and Electronics Conference (Naecon)*, pp. 246-249, 2012.
[142] M. I. Skolnik, "Introduction to radar systems," 3rd ed. New York, NY: McGraw-Hill, 2001.
[143] J. Capmany, B. Ortega, and D. Pastor, "A tutorial on microwave photonic filters," *Journal of Lightwave Technology,* vol. 24, no. 1, pp. 201-229, Jan, 2006.
[144] J. Capmany, B. Ortega, D. Pastor, and S. Sales, "Discrete-time optical processing of microwave signals," *Journal of Lightwave Technology,* vol. 23, no. 2, pp. 702-723, Feb, 2005.
[145] A. Ortigosa-Blanch, J. Mora, J. Capmany, B. Ortega, and D. Pastor, "Tunable radio-frequency photonic filter based on an actively mode-locked fiber laser," *Optics Letters,* vol. 31, no. 6, pp. 709-711, Mar 15, 2006.
[146] E. Hamidi, D. E. Leaird, and A. M. Weiner, "Tunable Programmable Microwave Photonic Filters Based on an Optical Frequency Comb," *Ieee Transactions on Microwave Theory and Techniques,* vol. 58, no. 11, pp. 3269-3278, Nov, 2010.
[147] X. X. Xue, Y. Xuan, H. J. Kim, J. Wang, D. E. Leaird, M. H. Qi, and A. M. Weiner, "Programmable Single-Bandpass Photonic RF Filter Based on Kerr Comb from a Microring," *Journal of Lightwave Technology,* vol. 32, no. 20, pp. 3557-3565, Oct 15, 2014.
[148] J. McClellan, T. W. Parks, and L. Rabiner, "A computer program for designing optimum FIR linear phase digital filters," *Ieee Transactions on Audio and Electroacoustics*, vol. 21, no. 6, pp. 506-526, Dec, 1973.
[149] J. Capmany, J. Mora, I. Gasulla, J. Sancho, J. Lloret, and S. Sales, "Microwave Photonic Signal Processing," *Journal of Lightwave Technology,* vol. 31, no. 4, pp. 571-586, Feb 15, 2013.
[150] W. L. Liu, M. Li, R. S. Guzzon, E. J. Norberg, J. S. Parker, M. Z. Lu, L. A. Coldren, and J. P. Yao, "A fully reconfigurable photonic integrated signal processor," *Nature Photonics,* vol. 10, no. 3, pp. 190-+, Mar, 2016.
[151] K. A. Rutkowska, D. Duchesne, M. J. Strain, R. Morandotti, M. Sorel, and J. Azana, "Ultrafast all-optical temporal differentiators based on CMOS-compatible integrated-waveguide Bragg gratings," *Optics Express,* vol. 19, no. 20, pp. 19514-19522, Sep 26, 2011.
[152] J. Y. Wu, P. Cao, X. F. Hu, X. H. Jiang, T. Pan, Y. X. Yang, C. Y. Qiu, C. Tremblay, and Y. K. Su, "Compact tunable silicon photonic differential-equation solver for general linear time-invariant systems," *Optics Express,* vol. 22, no. 21, pp. 26254-26264, Oct 20, 2014.
[153] X.Xu, et al., "Microwave Photonic All-optical Differentiator based on an Integrated Frequency Comb Source", Applied Physics Letters (APL) Photonics **2** 096104 (2017).
[154] M. R. E. Lamont, Y. Okawachi, and A. L. Gaeta, "Route to stabilized ultrabroadband microresonator-based frequency combs," *Optics Letters,* vol. 38, no. 18, pp. 3478-3481, Sep 15, 2013.
[155] J. Y. Wu, X. H. Jiang, T. Pan, P. Cao, L. Zhang, X. F. Hu, and Y. K. Su, "Non-blocking 2 x 2 switching unit based on nested silicon microring resonators with high extinction ratios and low crosstalks," *Chinese Science Bulletin,* vol. 59, no. 22, pp. 2702-2708, Aug, 2014.
[156] J. Y. Wu, P. Cao, T. Pan, Y. X. Yang, C. Y. Qiu, C. Tremblay, and Y. K. Su, "Compact on-chip 1 x 2 wavelength selective switch based on silicon microring resonator with nested pairs of subrings," *Photonics Research,* vol. 3, no. 1, pp. 9-14, Feb, 2015.
[157] T. Carmon, L. Yang, and K. J. Vahala, "Dynamical thermal behavior and thermal self-stability of microcavities," *Optics Express,* vol. 12, no. 20, pp. 4742-4750, Oct 4, 2004.
[158] T. Kobatake, T. Kato, H. Itobe, Y. Nakagawa, and T. Tanabe, "Thermal Effects on Kerr Comb Generation in a CaF2 Whispering-Gallery Mode Microcavity," *Ieee Photonics Journal,* vol. 8, no. 2, Apr, 2016.
[159] D. V. Strekalov, and N. Yu, "Generation of optical combs in a whispering gallery mode resonator from a bichromatic pump," *Physical Review A,* vol. 79, no. 4, Apr, 2009.
[160] T. Hansson, and S. Wabnitz, "Bichromatically pumped microresonator frequency combs," *Physical Review A,* vol. 90, no. 1, Jul 10, 2014.
[161] S. B. Papp, P. Del'Haye, and S. A. Diddams, "Parametric seeding of a microresonator optical frequency comb," *Optics Express,* vol. 21, no. 15, pp. 17615-17624, Jul 29, 2013.

[162] P. Del'Haye, K. Beha, S. B. Papp, and S. A. Diddams, "Self-Injection Locking and Phase-Locked States in Microresonator-Based Optical Frequency Combs," *Physical Review Letters,* vol. 112, no. 4, Jan 29, 2014.

[163] A. Pasquazi, L. Caspani, M. Peccianti, M. Clerici, M. Ferrera, L. Razzari, D. Duchesne, B. E. Little, S. T. Chu, D. J. Moss, and R. Morandotti, "Self-locked optical parametric oscillation in a CMOS compatible microring resonator: a route to robust optical frequency comb generation on a chip," *Optics Express,* vol. **21**, no. 11, pp. 13333-13341, (2013).

[164] C. Reimer, L. Caspani, M. Clerici, M. Ferrera, M. Kues, M. Peccianti, A. Pasquazi, L. Razzari, B. E. Little, S. T. Chu, D. J. Moss, and R. Morandotti, "Integrated frequency comb source of heralded single photons," *Optics Express,* vol. **22**, no. 6, pp. 6535-6546, (2014).

[165] C. Joshi, J. K. Jang, K. Luke, X. C. Ji, S. A. Miller, A. Klenner, Y. Okawachi, M. Lipson, and A. L. Gaeta, "Thermally controlled comb generation and soliton modelocking in microresonators," *Optics Letters,* vol. 41, no. 11, pp. 2565-2568, Jun 1, 2016.

[166] X. Yi, Q. F. Yang, K. Y. Yang, M. G. Suh, and K. Vahala, "Soliton frequency comb at microwave rates in a high-Q silica microresonator," *Optica,* vol. 2, no. 12, pp. 1078-1085, Dec 20, 2015.

[167] X. Yi, Q. F. Yang, K. Y. Yang, and K. Vahala, "Active capture and stabilization of temporal solitons in microresonators," *Optics Letters,* vol. 41, no. 9, pp. 2037-2040, May 1, 2016.

[168] P. Del'Haye, S. B. Papp, and S. A. Diddams, "Hybrid Electro-Optically Modulated Microcombs," *Physical Review Letters,* vol. 109, no. 26, Dec 26, 2012.

[169] W. Liang, V. S. Ilchenko, A. A. Savchenkov, A. B. Matsko, D. Seidel, and L. Maleki, "Whispering-gallery-mode-resonator-based ultranarrow linewidth external-cavity semiconductor laser," *Optics Letters,* vol. 35, no. 16, pp. 2822-2824, Aug 15, 2010.

[170] A. A. Savchenkov, D. Eliyahu, W. Liang, V. S. Ilchenko, J. Byrd, A. B. Matsko, D. Seidel, and L. Maleki, "Stabilization of a Kerr frequency comb oscillator," *Optics Letters,* vol. 38, no. 15, pp. 2636-2639, Aug 1, 2013.

[171] C.Reimer, et al., "Cross-polarized photon-pair generation and bi-chromatically pumped optical parametric oscillation on a chip", Nature Communications **6** Article 8236 (2015). DOI: 10.1038/ncomms9236

[172] T. Kessler, C. Hagemann, C. Grebing, T. Legero, U. Sterr, F. Riehle, M. J. Martin, L. Chen, and J. Ye, "A sub-40-mHz-linewidth laser based on a silicon single-crystal optical cavity," *Nature Photonics,* vol. **6**, no. 10, pp. 687-692, Oct, 2012.